# Computational framework for the generation of one-dimensional vascular models accounting for uncertainty in networks extracted from medical images

Michelle A. Bartolo[1,*], Alyssa M. Taylor-LaPole[1,*], Darsh Gandhi[1,4], AlexandriaJohnson[1,5], Yaqi Li[1,6], Emma Slack[1,7], Isaiah Stevens[1], Zachary Turner[1,8], Justin D. Weigand[2], Charles Puelz[2], Dirk Husmeier[3], and Mette S. Olufsen[1]

[1]*Department of Mathematics, North Carolina State University, Raleigh, North Carolina, USA*
[2]*Division of Cardiology, Department of Pediatrics, Baylor College of Medicine, Houston, TX, USA*
[3]*School of Mathematics and Statistics, University of Glasgow, Glasgow, UK*
[4]*Department of Mathematics, University of Texas at Arlington, Arlington, TX, USA*
[5]*Department of Mathematics and Statistics, University of South Florida, Tampa, FL, USA*
[6]*North Carolina School of Science and Mathematics, Durham, NC, USA*
[7]*Department of Mathematics, Colorado State University, Fort Collins, CO, USA*
[8]*School of Mathematical and Statistical Sciences, Arizona State University, Tempe, AZ, USA,*
[*]*Authors contributed equally*

## ABSTRACT

One-dimensional (1D) cardiovascular models offer a non-invasive method to answer medical questions, including predictions of wave-reflection, shear stress, functional flow reserve, vascular resistance, and compliance. This model type can predict patient-specific outcomes by solving 1D fluid dynamics equations in geometric networks extracted from medical images. However, the inherent uncertainty in in-vivo imaging introduces variability in network size and vessel dimensions, affecting hemodynamic predictions. Understanding the influence of variation in image-derived properties is essential to assess the fidelity of model predictions. Numerous programs exist to render three-dimensional surfaces and construct vessel centerlines. Still, there is no exact way to generate vascular trees from the centerlines while accounting for uncertainty in data. This study introduces an innovative framework employing statistical change point analysis to generate labeled trees that encode vessel dimensions and their associated uncertainty from medical images. To test this framework, we explore the impact of uncertainty in 1D hemodynamic predictions in a systemic and pulmonary arterial network. Simulations explore hemodynamic variations resulting from changes in vessel dimensions and segmentation; the latter is achieved by analyzing multiple segmentations of the same images. Results demonstrate the importance of accurately defining vessel radii and lengths when generating high-fidelity patient-specific hemodynamics models.

## KEY POINTS

- This study introduces novel algorithms for generating labeled directed trees from medical images, focusing on accurate junction node placement and radius extraction using change points to provide hemodynamic predictions with uncertainty within expected measurement error.
- Geometric features, such as vessel dimension (length and radius) and network size, significantly impact pressure and flow predictions in both pulmonary and aortic arterial networks.
- Standardizing networks to a consistent number of vessels is crucial for meaningful comparisons and decreases hemodynamic uncertainty.
- Change points are valuable to understanding structural transitions in vascular data, providing an automated and efficient way to detect shifts in vessel characteristics and ensure reliable extraction of representative vessel radii.

## INTRODUCTION

Medical images, including computed tomography (CT) and magnetic resonance imaging (MRI), are widely used for disease diagnostics and treatment planning (Chassidim et al., 2015; Doi, 2006; Khoo et al., 1997; Alasti et al., 2006). They are non-invasive and reliable ways to assess the structure of the heart, lungs, and vasculature. However, such images do not provide insight into hemodynamics, which is essential to determine the health of the cardiovascular system. One way to assess cardiovascular health is to embed one-dimensional (1D) patient-specific computational fluid dynamics (CFD) models within the image. This can be done by extracting a patient-specific vascular network from the medical image, and then solving fluid dynamics equations in each vessel within the network to make predictions of dynamic blood pressure and flow. To generate high-fidelity predictions for clinical applications, it is essential to minimize and quantify the uncertainty associated with this process. In this context, uncertainty arises from two main sources: the network extracted from the medical image and parameters in the computational model. Numerous studies have explored how variations in model parameters influence fluid dynamics predictions (Bertaglia et al., 2020; Chen et al., 2013; Ninos et al., 2021; Ye et al., 2022), but only a few (Colebank et al., 2019; Sankaran et al., 2015) have addressed the role of variations in network geometry. This study focuses on determining variations in vessel radius, length, and network connectivity and exploring how it affects 1D hemodynamic predictions in arterial networks.

Labeled trees are generated from three-dimensional (3D) rendered surfaces obtained using image segmentation. Image segmentation allows for analysis of anatomical data by isolating key features in an image, in this case vascular networks (Pham et al., 2000; Patil and Deore, 2013). However, there is inherent uncertainty in the image segmentation process due to complex anatomy, patient motion during scanning, variability in user expertise, and noise (Sharma and Aggarwal, 2010; O'Donnell, 2001; van Rikxoort and van Ginneken, 2013; Buelow et al., 2005). Despite advances including semi- and fully automatic extraction using machine learning, most software requires manual editing to capture the entirety of the desired anatomical features (van Rikxoort and van Ginneken, 2013). In particular, patient images are only captured up to a finite resolution, and certain anatomies, such as the rapidly branching structure of the pulmonary vasculature, are nearly impossible to extract in an automated manner (Colebank et al., 2019; van Rikxoort and van Ginneken, 2013). Other anatomies, such as the aorta, are difficult to automatically extract due to differences in spatial scales within the structures we wish to capture, which includes the smaller branching neck vessels and larger main aortic branch. Overall, a major challenge is that vascular networks span several orders of magnitude, but resolution within the image is constant (Lakhani, 2020), adding significant uncertainty when extracting data, especially for small vessels (Colebank et al., 2019). Moreover, for images acquired without contrast, differentiating arteries, veins, and other structures is not trivial (van Rikxoort and van Ginneken, 2013). This study generates networks from images segmented using 3D Slicer developed by Kitware, Inc. (Federov et al., 2012), which is a widely used open-source software.

After segmenting the image and generating a 3D rendered surface, centerlines are identified along each vessel, creating a 1D characterization of blood vessel geometry (Pham et al., 2000; Patil and Deore, 2013). This study builds networks using the Vascular Modeling Toolkit (VMTK) (Izzo et al., 2018), which generates centerlines by placing maximally inscribed spheres along each vessel. This process is sensitive to intricate vascular networks containing vessels with high tortuosity, rapid branching structures, and narrow lumen. In these networks, maximally inscribed spheres placed by VMTK intersect earlier than expected, leading to junction nodes placed outside of physiological domain. In particular, VMTK often positions junction nodes prior to the *ostium region*, the area surrounding the opening of a blood vessel where it divides into multiple vessels and where the radius is not defined (Cheng, 2019). This limitation has been described in several studies, including Ellwein et al. (2016) and Pfaller et al. (2022). To address the limitation, Ellwein et al. (2016) adjust junctions by examining the angles between vessels, while Pfaller et al. (2022) utilize slices rather than maximally inscribed spheres to generate centerlines. Other programs exist to generate centerlines, such as SimVascular (Updegrove et al., 2017), CRIMSON (CardiovasculaR Integrated Modelling and SimulatiON) (Arthurs et al., 2021), and SGEXT (Cappetti et al., 2022), each balancing accuracy and efficiency. Independent of the software used to generate centerlines, the vessel radius is not defined in the ostium region, yet fluid dynamics models solving the equations in 1D trees require a radius measurement at all points along the vessel. To address this limitation, this study devises an algorithm to move junctions into the ostium region and uses statistical change points to identify a segment within each vessel's centerlines that best represents its radius.

CFD modeling provides insights into the function of the vasculature by predicting hemodynamic properties that cannot be measured in vivo (Mittal et al., 2016; Chinnaiyan et al., 2017). By integrating with patient-specific networks, these methods can be used to study the outcome of treatments or predict disease progression (Candreva et al., 2022). Several recent studies have used patient-specific geometry derived from medical images to predict hemodynamics (Taylor-LaPole et al., 2023; Colebank et al., 2019; Bartolo et al., 2022; Battista et al., 2016; Formaggia et al., 2006; Gray and Pathmanathan, 2018; Yang et al., 2019). They superimpose dynamic pressure and or flow data onto geometric domains extracted from images and make predictions of blood pressure, flow, shear stress, and wave intensity to simulate the impact of disease and treatment. The most detailed models include 3D CFD and fluid-structure interaction, which provide predictions of local flow patterns important in regions with significant secondary flows. Unfortunately, 3D models are computationally expensive and challenging to calibrate to patient data (Yang et al., 2019; Botnar et al., 2000). Thus, this study focuses on 1D fluid dynamics models, providing an efficient alternative to predict volumetric flow and pressure averaged over vessel cross-sections. This model type is useful for predicting wave-propagation and wave-intensity, organ perfusion, functional flow reserve, vascular resistance, and compliance (Shi et al., 2011). In addition, 1D models can be easily calibrated to data, making them ideal for patient-specific simulations, and provide reasonable agreement to 3D models (Moore et al., 2005; Reymond et al., 2012; Blanco et al., 2018).

Uncertainty arises in nearly every level of the modeling process, including in the assumptions to derive the equations that model the physical system, parameter choices, and geometry extracted from image segmentation. Uncertainty quantification has been studied extensively in computational modeling of hemodynamics. For example, Colebank et al. (2019) study the influence of pre-segmentation parameters, such as thresholding and smoothing, on 1D predictions in ex vivo murine arteries. Chen et al. (2013) incorporates intrinsic parametric uncertainties to depict blood flow and pressure within a stochastic model of arteries. Bertaglia et al. (2020) highlight the effects of changes in elastic and viscoelastic parameters in fluid-structure interaction models of arteries. While these studies effectively quantify uncertainty in the modeling and parameter estimation process, they do not examine uncertainties associated with the network extraction process. It is well known that there is a significant interplay between vascular geometry and hemodynamics, and therefore, the uncertainty of geometric features of vessels must be explored and quantified (Gounley et al., 2017).

This study investigates the process of extracting networks from medical images and the associated hemodynamic uncertainty. We use 3D Slicer (Federov et al., 2012) to generate 3D rendered surfaces from CT images of the pulmonary arteries for a normotensive human subject and from magnetic resonance angiography (MRA) from a double outlet right ventricle (DORV) patient. VMTK is used to generate centerlines along each vessel. From these, we generate labeled directed trees characterized by vessel connectivity, radius, and length. To improve fidelity of the generated models, we devise algorithms to align networks and adjust junctions to physiologically accurate locations. We then employ statistical change point analysis to determine a representative radius for each vessel, avoiding outlier data. We propagate uncertainty in segmentation, junction location, and radius to a 1D fluid dynamics model, finding that variation in vessel radius have the most significant impact on fluid dynamics predictions. In addition, we generate multiple segmentations from each image – five for the pulmonary vasculature and ten for the aorta – to study variations between segmentations. Finally, in the pulmonary vasculature, each segmentation captures a varying number of vessels, so to enable comparison, we standardize networks using a radius pruning algorithm. Our framework generates high-fidelity labeled trees and provides hemodynamic predictions with uncertainty within expected measurement error. As patient-specific modeling becomes increasingly popular in medical applications, we argue that quantifying geometric uncertainty using our methodology (illustrated in Figure 1) is critical to obtaining reliable hemodynamic predictions.

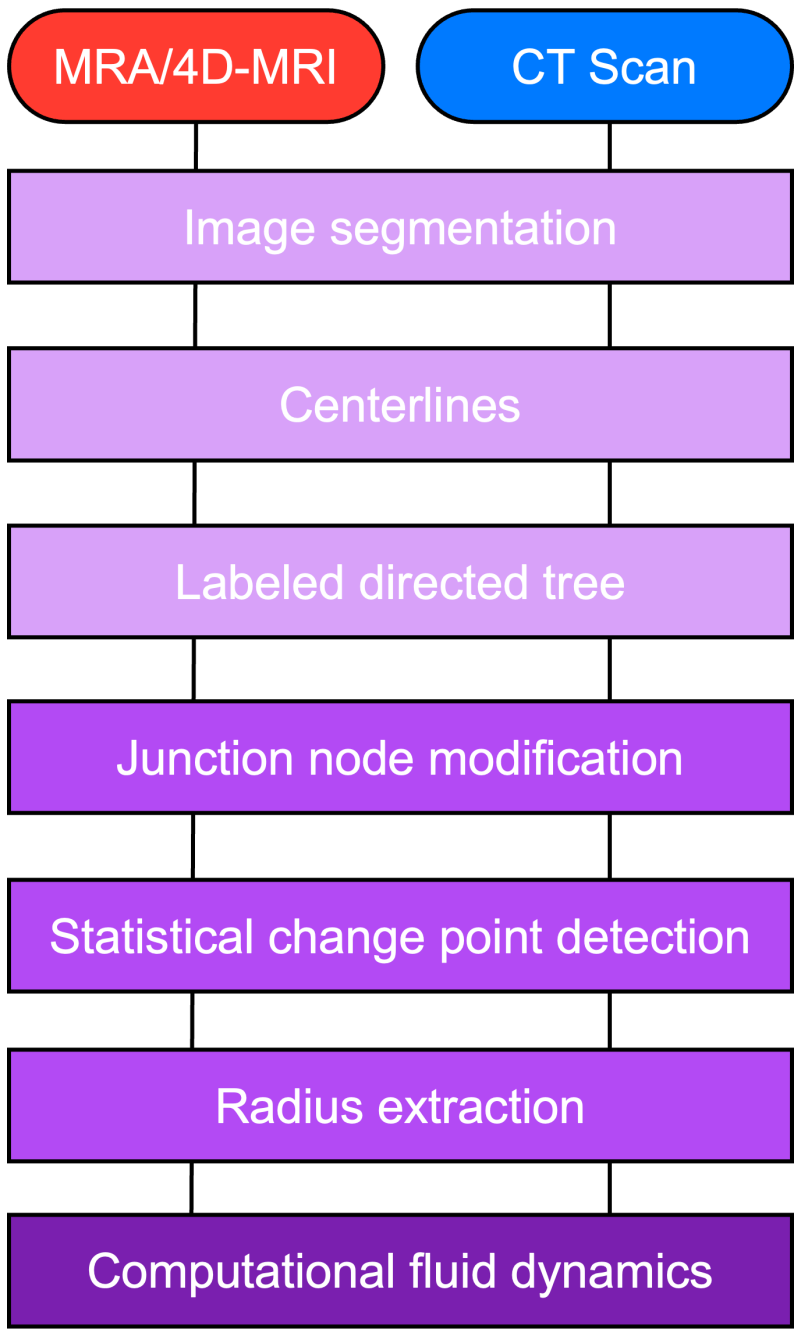

Figure 1: Workflow for our methodology, consisting of image segmentation and analysis, creation of labeled directed tree from centerlines, junction node modification, change point analysis, radius identification, and construction of a computational domain for fluid dynamics simulations.

# MATERIALS AND METHODS

## Image Analysis

We analyze two images: a chest CT image from a healthy, 67-year-old female volunteer captured using the Omnipaque 350 contrast agent available at the Vascular Model Repository (Wilson et al., 2013) (http://simvascular.github.io/) and an MRA image from a DORV patient acquired at the Texas Children's Hospital Heart Center. Data collection was approved by the Baylor College of Medicine, Institutional Review Board (H-46224: "Four-Dimensional Flow Cardiovascular Magnetic Resonance for the Assessment of Aortic Arch Properties in Single Ventricle Patients"). The MRA image was obtained using time-resolved, contrast-enhanced imaging performed using localizing sequences from the cardiac MRI with a 0.1 mL/kg intravenous gadolinium contrast bolus injection. More details on the image analysis protocol can be found in (Taylor-LaPole et al., 2023).

**3D Rendering.** The pulmonary and aortic networks are rendered as a 3D surface using the open-source image segmentation software 3D Slicer developed by Kitware, Inc. (see http://www.slicer.org) (Federov et al., 2012; Kikinis et al., 2014). Image intensities in the CT image range from 50 – 3027 Hounsfield unit (HU) and from $75 - 327$ HU for the MRA. Image voxel size for the CT image is $0.68 \times 0.68 \times 5.00$ mm$^3$ and is $1.22 \times 1.22 \times 1.40$ mm$^3$ for the MRA. The CT image was segmented five times and the MRA image 10 times by the same user, each using the same threshold and smoothing factors. Example 3D rendered surfaces for the two geometries are shown in Figure 2.

The MRA segmentation was acquired utilizing 3D Slicer's thresholding, growing from seeds, cutting, and smoothing tools in a semi-automatic manner. To segment pulmonary arterial networks from the CT image, we first identify the main (MPA), right (RPA), and left (LPA) pulmonary arteries through thresholding followed by manual painting, erasing, and cutting. This is followed by manual segmentation of the lobar, segmental, and subsegmental vessels (Colebank et al., 2021b). The rendered surfaces are saved as standard tessellation language (STL) files, which can be opened in Paraview, Kitware Inc. (Ayachit, 2015). The skeletonized images are converted to a VTK polygonal data file and then processed through VMTK (Antiga et al., 2008) (see http://www.vmtk.org).

**Centerlines.** VMTK (Izzo et al., 2018) determines centerlines in the 3D rendered surfaces of arteries using maximally inscribed spheres. Spheres, which touch at least four points on the 3D rendering, are inscribed throughout each vessel, and the medial axis is approximated by a Voronoi diagram (Antiga et al., 2008). Centerlines are defined as the minimal path along the inverse of the spheres' radii (Antiga et al., 2008). User-specified inlet and outlet points establish centerline boundaries, guiding VMTK in a recursive process starting at the terminal vessels. When two centerlines intersect, a junction node is placed. Example 3D rendered surfaces with centerlines are shown in Figure 2.

## Labeled Tree Generation

A tree is generated to form arterial networks and used in fluid dynamics simulations. This tree consists of two parts: (a) large arteries identified from centerlines derived from MRI and CT images and (b) small vessels represented by fractal trees. The small vessels extend the networks from (a) representing vessels that are not visible in the images. The caliber of vessels represented by fractal trees will depend on the image resolution.

Custom MATLAB algorithms are used to construct a raw labeled directed tree from VMTK centerlines (Colebank et al., 2019, 2021a). This algorithm transforms centerlines to edges with $x,y, \text{ and }z$ coordinates at the center of each maximally inscribed sphere and defines junctions as nodes where two or more centerlines intersect. The tree has three types of vessels: the *root vessel*, which starts at the inlet and ends at the first junction node, *central vessels*, which begin and end at junction nodes, and *terminal vessels*, which start at a junction node and end at an outlet node. Each vessel contains $n$ nodes with associated $x, y$, and $z$ values, radius, and length. The radii along the vessel are determined from the maximally inscribed spheres. Each vessel is assigned a length specified by calculating the Euclidean distances between $x, y$, and $z$ coordinates of all nodes within the vessel. Information about each vessel is stored in a connectivity matrix that links parent and daughter vessels to describe how vessels are connected in the tree.

The ten aortic networks all have 13 bifurcating vessels joined using the same connectivity matrix, but vessel length and radii vary between networks. In contrast, the rapidly branching pulmonary network requires manual segmentation to separate arteries from veins and occasionally leads to trifurcations. Therefore, in addition to vessel radius and length variation, the five segmented pulmonary arterial networks have different numbers of vessels and connectivity. The trees range from 167 to 238 vessels with $0.5 - 2\%$ of parents having three daughters and the rest of the parents having two daughters. There are no quadfurcations or higher in the networks we segment. To compare these networks, similar to our previous study (Miller et al., 2023) we employ a pruning procedure to generate trees with the same number of vessels. This involves iteratively removing the smallest terminal vessels with terminal siblings until the networks reach the desired number of vessels (Figure 3). In this case, we prune networks until they reach the size of the smallest network of 167 vessels. As seen in Figure 3, distinct networks have a more similar radius distribution after pruning.

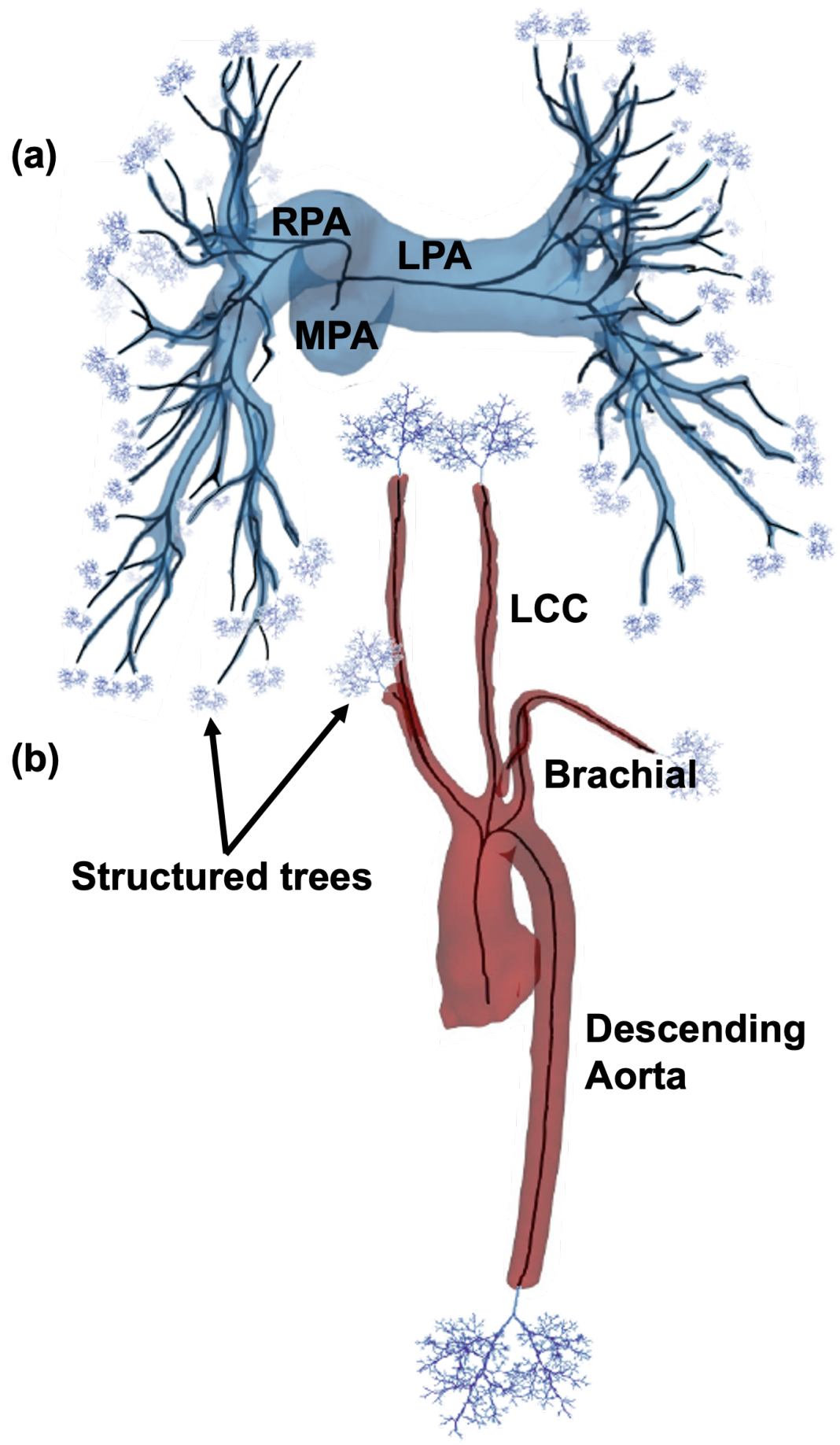

Figure 2: 3D rendering from (a) a normotensive pulmonary arterial network and (b) the aortic vasculature of a double outlet right ventricle patient. Centerlines (black lines) and structured trees, used outside the imaged region, constitute a labeled directed tree.

**Small Vessels.** Due to limits in image resolution, it is not possible to obtain measurements for small arteries and arterioles. To represent all arteries in the network, starting at the heart and ending at the capillary level (Boron and Boulpaep, 2017), we augment the large vessel domain with asymmetrically-branching, self-similar fractal trees (Olufsen et al., 2000). These structured trees have a root vessel with the same radius as the large terminal vessel it is attached to. Each parent vessel, $p$, branches into two daughter vessels $\boldsymbol{d} = (d_1, d_2)$, with length and radii defined as

$$r_{d1} = \alpha r_p, \qquad r_{d2} = \beta r_p, \qquad l_r = r_d l_{rr} \tag{1}$$

where $\alpha, \beta < 1, \alpha + \beta \geq 1$, and $\ell_{rr}$ is a length to radius ratio. Values for these parameters are listed in Table 1 and are assigned based on literature and analysis of murine micro-CT images (Olufsen et al., 2000; Chambers et al., 2020). The structured trees bifurcate until a specified minimum radius, $r_{\min}$, is reached. For this study, $r_{\min} = 0.001$ cm, consistent with a red blood cell's diameter (Townsley, 2012).

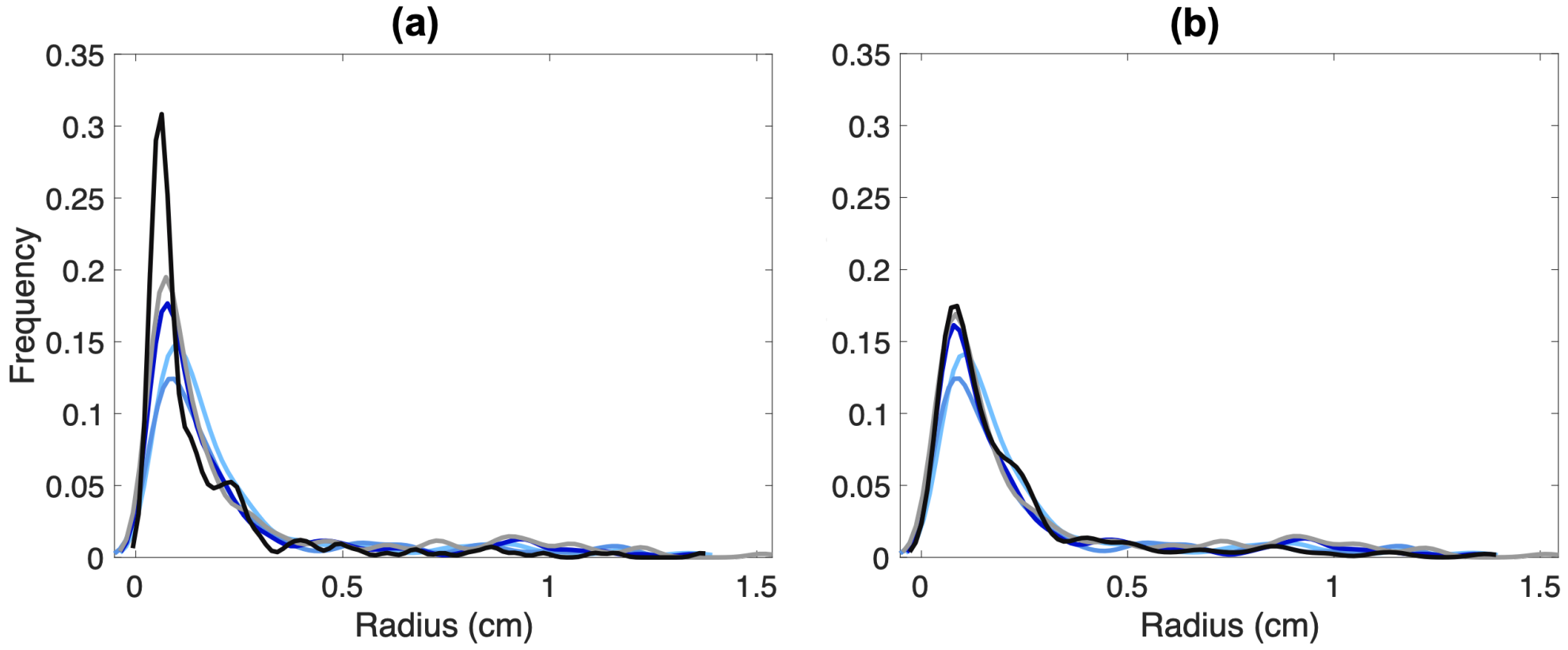

Figure 3: Kernel density estimates in five pulmonary arterial networks for trees (a) before and (b) after pruning to have the same number of vessels.

## Adjusted Labeled Trees

The directed trees representing the large vessels have two flaws: (1) the junction nodes are often placed outside of the ostium, and (2) the low-dimensional representation of the vessel radius is incorrectly estimated if all nodes within a vessel are used. In particular, the radii assigned for vessel nodes within the ostium do not represent the vessel's true radius, as it contains values for multiple vessels as they split. To mitigate these limitations, we have designed an algorithm that moves the junction node to the center of the ostium and identifies the segment within each vessel that best represents its radius. Nodes along this optimal segment are then used to determine a radius and its standard deviation along the entire vessel. The input to our algorithm is a connectivity matrix, vessel length and radii generated by VMTK, and post-processed by custom algorithms devised in this study. Results include an adjusted labeled tree that reduce the uncertainty in fluid dynamics predictions propagated from uncertainty in vessel radii and length.

### Adjusted Junction Nodes

**Step 1:** Starting at the VMTK-derived junction nodes $\boldsymbol{x}_j = (x_j, y_j, z_j)$ that connect terminal vessels, we compute the distance between the $x, y, z$ coordinates of the daughter vessels as

$$D_i = \sqrt{\left(x_{d_1,i} - x_{d_2,i}\right)^2 + \left(y_{d_1,i} - y_{d_2,i}\right)^2 + \left(z_{d_1,i} - z_{d_2,i}\right)^2}, \qquad i = 1, \dots, N, \tag{2}$$

where $x_{d,i} = (x_{d,i}, y_{d,i}, z_{d,i})$ denotes the $i$th point after the junction node $\boldsymbol{x}_j$ for each of the daughter vessels $\boldsymbol{d} = (d_1, d_2, \dots, d_\#)$, where # is the number of daughter vessels and $x_{d,N}$ denotes the last node of the shortest daughter vessel. If there is a trifucation, we consider the pair of daughters that has the smallest distance between them. For each node $x_{d,i}$, we then normalize $D_i$ by dividing by the maximum distance between the daughter vessels, $D_{\max}$. A new junction node $\tilde{x}_j$ is placed at the smallest $i$ for which $D_i/D_{\max} > 0.1$. The cutoff value 0.1 is an empirically chosen input parameter. Figure 4(a) shows the $j$th junction node representing the end of the parent vessel that separates into a bifurcation and nodes along each of the daughter vessels $\boldsymbol{x}_{d,i}$, where $d_1$ is red and $d_2$ is blue.

**Step 2:** Starting at the first node along each daughter vessel $\boldsymbol{x}_{d,1}$, we identify all nodes $i = 1, \dots, k$ for which $D_i/D_{\max} \leq 0.1$. Here, $k$ denotes the index of first the node for which $D_i/D_{\max} > 0.1$. The new nodes along the parent branch are computed by averaging the location of the daughter vessel coordinates

$$x_{p,i} = \frac{1}{\#}\sum_{\ell=1}^{\#} x_{d_\ell,i}. \tag{3}$$

Figure 4(b) shows the new parent vessel nodes (green) and the adjusted junction node $\tilde{\boldsymbol{x}}_j$ (purple). Note that the path from the new junction to the daughter vessels is not smooth.

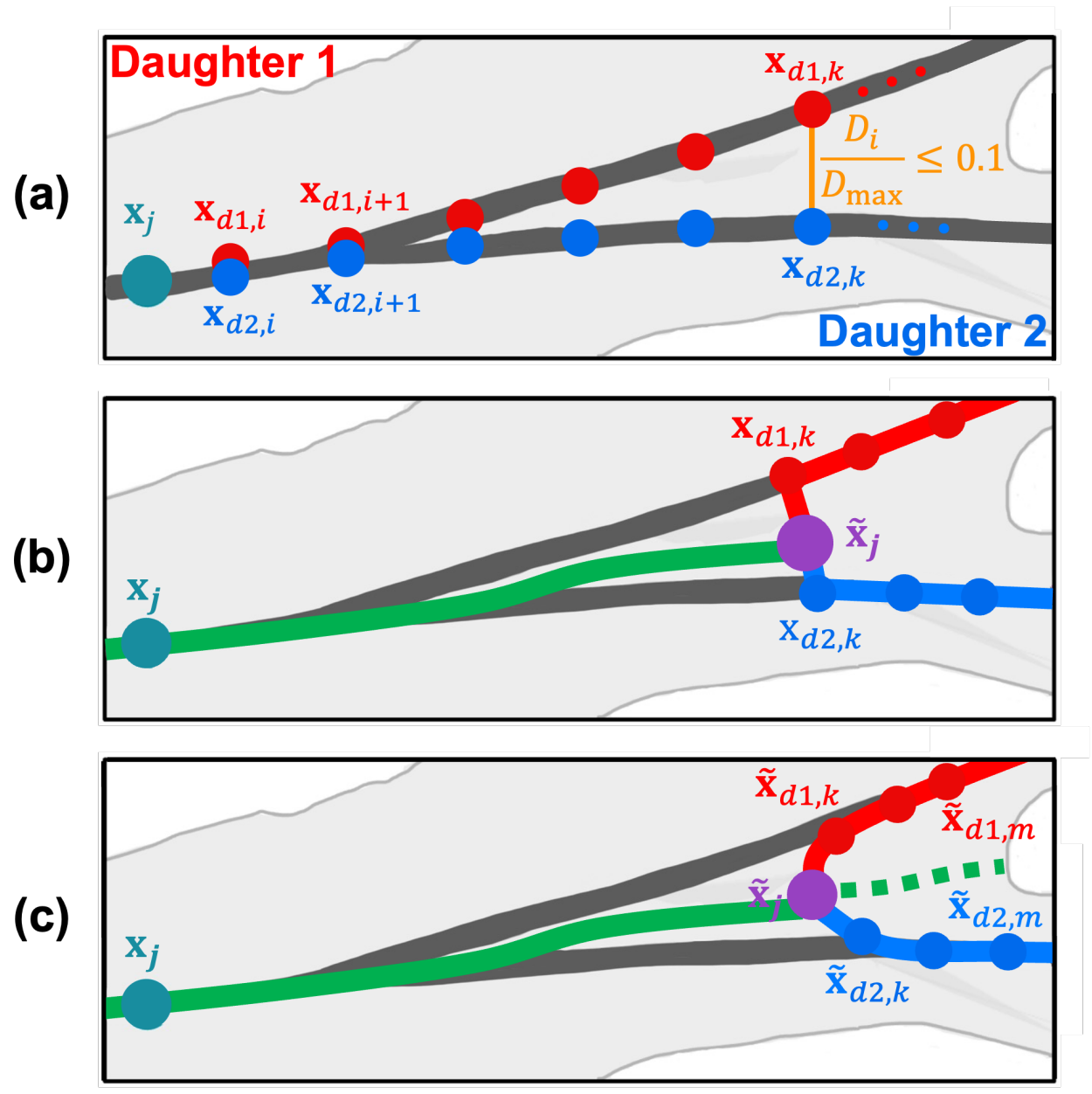

Figure 4: Junction node adjustment in 3D space for a bifurcating vessel. Dark gray lines and teal circle represent the VMTK-generated centerlines and junction node $\boldsymbol{x}_j$, respectively. (a) Nodes along daughter vessels, where daughter vessel 1, $d_1$, is in red and daughter vessel 2, $d_2$, is in blue, for which $D_i/D_{max} \leq 0.1, i = 1, \dots, k$. (b) The green line and purple circle denote the new parent vessel and junction node $\tilde{\boldsymbol{x}}_j$, calculated by an average between daughter

vessel nodes. The direct path from the new junction nodes to daughter vessels $x_{d,k}$ is characterized by a sharp angle. (c) Smoothed adjustment of daughter nodes $\widetilde{\boldsymbol{x}}_{\boldsymbol{d},i}$, based on a weighted averaged in sections where $0.1 < D_i/D_{max} \leq 0.2,\ i = k, \dots, m$.

**Step 3:** The new junction node $\widetilde{\boldsymbol{x}}_j$ is connected to nodes $\boldsymbol{x}_{\boldsymbol{d},i}, i = 1, \dots, k$ along each daughter vessel. To ensure a smooth transition, we compute a weighted average in sections where daughter vessel centerlines are within $10 - 20\%$, i.e., nodes for which $0.1 < D_i/D_{\max} \leq 0.2, i = k, \dots, m$. Here, $m$ is the index of the first node for which $D_i/D_{\max} \geq 0.2$. The new daughter nodes, $\widetilde{\boldsymbol{x}}_{\boldsymbol{d},i}, i = 1, \dots, k$, shown in Figure 4(c), are computed as a weighted average between the extension of the parent branch (dashed green line on Figure 4(c)) and the original nodes $\widetilde{\boldsymbol{x}}_{\boldsymbol{d},i}, i = k, \dots, m$. The extension of the parent branch is computed by averaging all nodes $i = k, \dots, N$ along the daughter vessels as described in Equation (3).

The degree of weighting between the extended parent vessel and the daughter vessels is computed based on the ratio of the radius and the distance between the extended parent and the daughter vessels. The locations of the new daughter nodes are calculated as

$$\widetilde{\boldsymbol{x}}_{\boldsymbol{d},i} = \frac{w\boldsymbol{x}_{\boldsymbol{d},i} + \boldsymbol{x}_{p,i}}{w + 1}, i = k, \dots, m. \tag{4}$$

Each daughter is individually averaged with the extended parent using weights, i.e, $d_1, d_2, \dots, d_\#$ are averaged separately. For the pulmonary arteries, we use a double-weighted average ($w = 2$). Since the aorta consists of a large, curved main vessel, centerlines lie closer to the main aortic vessel than the neck vessels, branching upward at a nearly $90$ −degree angle. For these vessels, we calculate a $100$ −times weighted average $w = 100$ towards the main aortic branch. The weighting cutoff of $0.2$ and the degree of weighting $w$ are empirically chosen input parameters that can be adjusted to fit the particular application. Based on these adjustments, new lengths and radii are assigned to the labeled directed tree.

**Adjusted Radii**

The radii identified by VMTK at each node along vessels include values in the ostium region, oscillations associated with irregularities on the surface of the 3D rendering, and regions close to image resolution. Therefore, to determine the radius of each vessel, we devise a three-step process that (1) uses change points to characterize the vessel shape, (2) determines the region that represents the vessel radius (Algorithms 1 and 2), and (3) checks the consistency of the generated labeled tree.

**Step 1: Statistical change point calculation.** A change point $\psi \in \{1, \dots, n-1\}$ can be identified within the ordered sequence of radii data (determined from VMTK values) $\boldsymbol{r} = \{r_1, \dots, r_n\}$ for which statistical properties $\{r_1, \dots, r_\psi\}$ and $\{r_{\psi+1}, \dots, r_n\}$ differ significantly. The dataset may contain multiple change points, represented as $\boldsymbol{\psi} = \{\psi_1, \psi_2, \dots, \psi_m\}$, where $m$ is the number of change points. We use a two-step process to determine the change points using publicly available R packages. We use R/changepoint to identify the number of change points and R/segmented to define the location of change points within each vessel. The two-step process is needed since R/segmented requires user-specification of the number of change points, and R/changepoint does not precisely indicate where change points occur.

***Number of change points.*** We use the R/changepoint package (Killick and Eckley, 2014) to determine the optimal number of change points in each vessel. This program employs binary segmentation, iteratively partitioning the dataset and fitting a piecewise constant line in each segment. To determine if a change has occurred, the algorithm seeks the location at which the maximum of the log-likelihood is attained,

$$ML(\psi_1) = \log\left(p(\boldsymbol{r}_{1:\psi_1}|\hat{\theta}_1)\right) + \log\left(p(\boldsymbol{r}_{(\psi_1+1):n}|\hat{\theta}_2)\right), \tag{5}$$

where $\hat{\theta}_1$ and $\hat{\theta}_2$ are estimates of mean and variance before and after the data is partitioned. We use Bayesian Information Criterion ($BIC$) to penalize the $ML$

$$ML(\psi_1) = \log\left(p(\boldsymbol{r}_{1:\psi_1}|\hat{\theta}_1)\right) + \log\left(p(\boldsymbol{r}_{(\psi_1+1):n}|\hat{\theta}_2)\right), \tag{6}$$

where $\kappa$ is the number of parameters and $n$ is the number of data points. Here, $\kappa = 2$ since parameters are the mean and variance. Once the $BIC$ is minimized, a change point is placed at $\max_{\psi_1}\left(ML(\psi_1)\right)$. We limit the number of change points to three to avoid overfitting and enable consistent analysis between vessels. This binary change point algorithm efficiently identifies the number of change points at a low computational cost but at the expense of precise change point location (Eckley et al., 2011; Killick et al., 2012; Killick and Eckley, 2014), so we employ R/segmented to improve change point placement.

***Change point location.*** Optimal change point placement is determined using the customizable and flexible R/segmented package, which requires specification of the number of change points (selected as described above). Like R/changepoint, this program employs binary segmentation, iteratively partitioning the dataset. Within each segment, the algorithm maximizes the likelihood between a linear regression model and the data (Muggeo, 2003, 2008) (Figure 5), using a statistical model that accounts for the slope, mean, and variance, which are parameters of the regression model.

A two-step process is employed using a regression model between the mean response, $E[\boldsymbol{r}]$, and parameters, $\theta$, to determine the location of change points, and then a simpler regression model (using the same response and a predictor) to fit piecewise linear functions between change points. The regression model used to determine change point location is defined as

$$g(E[\boldsymbol{r}]) = \eta(l) + b \times h(\hat{\theta}; \psi_i), \tag{6}$$

where $g(\cdot)$ is the link function for $E[\boldsymbol{r}], h(\hat{\theta}; \psi_i)$ is a parameterization for $\theta, \psi_i, i = 1, 2, 3$ are the change points, $\eta(\boldsymbol{l})$ is the predictor, $\boldsymbol{l}$ is the length along the vessel. A first-order Taylor expansion around an initial $\psi_i^{(0)}$ is used to approximate

$$h(\hat{\theta}; \psi_i) \approx h(\hat{\theta}; \psi_i^{(0)}) + (\psi_i - \psi_i^{(0)})h'(\hat{\theta}; \psi_i^{(0)}). \tag{7}$$

The response and parameters form a piecewise relationship with the terms

$$a\theta + b(\theta - \psi_i)_+ \tag{8}$$

Note that $(\theta - \psi_i)_+ = (\theta - \psi_i) \times \left(I(\theta > \psi_i)\right)$ where $I(\cdot)$ is the indicator function. $I(\cdot) = 1$ when the statement is true, and $I(\cdot) = 0$, otherwise. Using this parameterization

- $a$ is the slope of the line segment prior to the change point.
- $b$ is the difference in slopes of the lines before and after the change point.
- $(a + b)$ is the slope of the line segment following the change point.

The change point is fixed at $U^{(s)} = \left(\theta - \psi_i^{(s)}\right)_+$ and $V^{(s)} = -I(\theta) > \psi_i^{(s)}$ and iterated $s$ times until there is no improvement in the $ML$ estimate. Using radii, $\boldsymbol{r}$ (cm), as the response, and $x_{\boldsymbol{d},i} = (1, \dots, k)$ the location along the vessel, the linear regression model is given by

$$\boldsymbol{r}(i) = \beta_0 - \beta_1 x_{\boldsymbol{d},i}, \tag{9}$$

where $\beta_0, \beta_1$ are regression coefficients.

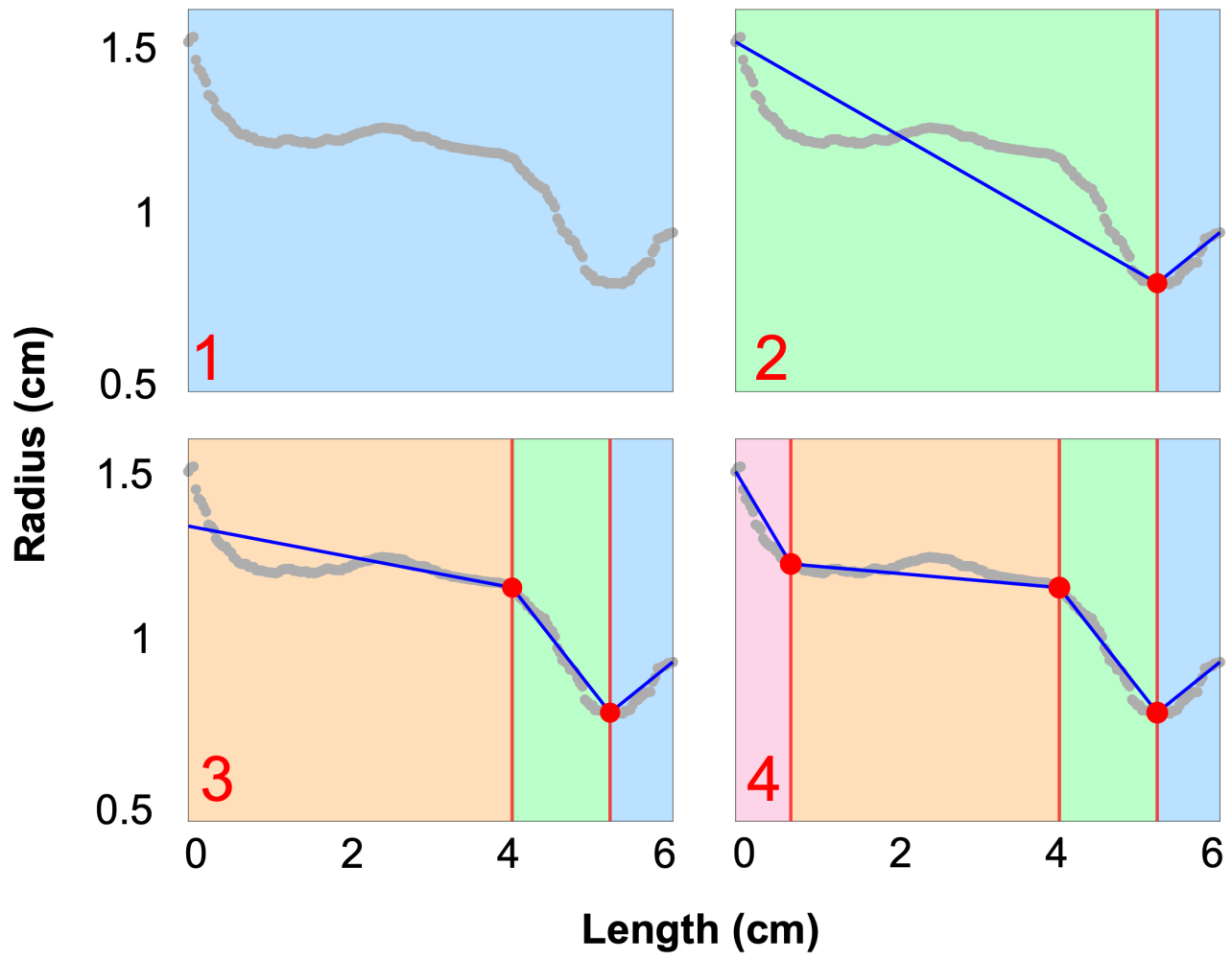

Figure 5: Binary segmentation is used to place change points along radii data attained from VMTK (grey points). In each panel, the blue lines are the piecewise linear fit determined in Equation (9), and the red circles denote the change point placement. Panel 1 shows the radii before change points are placed, and panel 2 shows the segmentation after placing the first change point. Panel 3 shows the placement of the second change point. It is placed to the left of the first change point as this placement leads to a higher maximum likelihood. Finally, panel 4 shows the optimal placement of three change points.

**Step 2: Vessel segment selection to adjust vessel radius and determine its uncertainty.** This step determines the segment within each vessel that best represents the vessel radius and from which the value can be reliably estimated. We refer to this as the optimal radius segment, $\boldsymbol{r}_{\text{opt}}$. Given the variation in vessel morphology, a multistep process is used to identify $\boldsymbol{r}_{\text{opt}}$. The process is described in detail in Algorithms 1 and 2 with R and MATLAB code available in the repository found at

https://github.com/msolufse/CDG_NCSU/tree/master/VascularTreeFromImages.

All vessels identified by VMTK contain $n$ nodes. Each node $x_i, i = 1, \dots, n$ is associated with a radius value $r(i)$. In step 1, each vessel is assigned between zero and three change points placed at $\psi_i, i = 1, \dots, m$, $m \leq 3$. Sections of nodes between change points are labeled by $\mathbf{sec}_i, i = 1, \ldots, m+1$. The first segment, $\mathbf{sec}_1 = \boldsymbol{r}(1, \psi_1)$, includes radii values from the first node of the vessel, $r(1)$, to the first change point $r(\psi_1)$. The last segment $\mathbf{sec}_{m+1} = \boldsymbol{r}(\psi_m, n)$ includes radii values from the last identified change point, $r(\psi_m)$, to the last node in the vessel, $r(n)$. The linear fits within each segment, also determined in step 1, are used to determine the slope of each segment $slp_i, i = 1, \ldots, m+1$.

Most 1D fluid dynamics models predict hemodynamics, assuming that vessels are either straight (Colebank et al., 2021a) or tapering (Olufsen et al., 2000). Typically, in rapidly branching vascular networks such as pulmonary vasculature, the brain, or the liver, vessels do not taper significantly between junctions, whereas in networks with longer vessels, e.g., networks including the aorta, carotid, brachial, or iliac arteries, vessels taper along their length (Taylor-LaPole et al., 2023; Olufsen et al., 2000; Epstein et al., 2015). The fluid dynamics code used in this study allows vessels to be straight or taper exponentially. Therefore, the proposed algorithm generates radii labels for both cases.

Non-tapering vessels are assigned a constant radius along the vessel, averaging values in $\boldsymbol{r}_{\text{opt}}$, such that

$$\hat{r} = \frac{1}{k}\sum_{i=1}^{k} r_{\text{opt}}(i), \tag{10}$$

where $r_{\text{opt}}(i)$ refers to the $i$th radius observation in $\boldsymbol{r}_{\text{opt}}$ and $k$ is the number of points in $\boldsymbol{r}_{\text{opt}}$. Note that although $\hat{\boldsymbol{r}}$ is assigned a single value based on the mean, it is a vector because we assign this value to every node along the length of the vessel. In tapering vessels, we fit a decaying exponential function to radius values in $\boldsymbol{r}_{\text{opt}}$ given by

$$\hat{\boldsymbol{r}}(i) = r_{\text{in}} \exp(i \log(r_{\text{in}}/r_{\text{out}})), \tag{11}$$

where $r_{\text{in}}$ and $r_{\text{out}}$ refer to inlet and outlet radius of the vessel. The parameters $r_{\text{in}}$ and $r_{\text{out}}$ are estimated using MATLAB's optimization function fminsearch() (MathWorks, 2022) minimizing the least squared error

$$J = \frac{1}{k}\sum_{i=1}^{k} \left(\hat{r}(i) - r_{\text{opt}}(i)\right)^2 \tag{12}$$

For all vessels, the standard deviation of the radius is determined by

$$\sigma_{\hat{r}} = \sqrt{\frac{\sum_{i=1}^{k}\left(r_{\text{opt}}(i) - \hat{r}(i)\right)}{k}}. \tag{13}$$

The following sections describe how $\boldsymbol{r}_{\text{opt}}$ is selected for the two vessel types.

***Non-tapering vessels.*** To identify $\boldsymbol{r}_{\text{opt}}$ for non-tapering vessels, the slopes of regression lines fit to $sec_i$ are considered, choosing the section with the smallest absolute value slope, $sec_{\min}$. The smallest slope is found as $slp_{\min} = \min(slp_i), i = 2, \ldots, m+1$. Note that the first segment (the section prior to the first change point) is not included as it is typically within the ostium region. To account for the variation in vessel morphometry, this process requires user specification of three parameters: the minimum percentage needed to ensure the length of single segment has sufficient data (here set at $n/4 = 25\%$), the threshold for analysis of multiple segments to specify $\boldsymbol{r}_{\text{opt}}$ (here set to $n/2 = 50\%$), and a threshold $\xi$ to compare the slopes of the last two segments (here set to $\xi = 2.8$).

*<u>Case 1</u>:* Identify the segment with the smallest absolute value slope located between change points $\psi_i, i = 1, \dots, m$ and $n$, the last node in the vessel. For a vessel with three change points, there are three sections to consider, $\boldsymbol{r}(\psi_1, \psi_2), \boldsymbol{r}(\psi_2, \psi_3)$, or $\boldsymbol{r}(\psi_3, n)$. If the section with the smallest absolute value slope includes more than 25% of the total vessel observations, we assign it to be $\boldsymbol{r}_{\text{opt}}$. The threshold $n/4 = 25\%$ is one of the three input parameters.

*<u>Case 2</u>:* For vessels not included in case 1, we determine the first ($i = 1$) and last ($i = k$) radius values in $\boldsymbol{r}_{\text{opt}}(i)$ by analyzing change point locations.

(a) To identify the first point in $\boldsymbol{r}_{\text{opt}}$, we consider the location of the first change point, $\psi_1$. If $\psi_1$ is position beyond the 25% mark in the vessel, $\boldsymbol{r}_{\text{opt}}$ begins at $r(\psi_1)$. If $\psi_1$ is located before the 25% mark, the first change point is assumed to be in the ostium region, so we start $\boldsymbol{r}_{\text{opt}}$ at the second change point, $r(\psi_2)$. To avoid excluding data, when $\psi_2$ is located more than 50% into the vessel, we instead start $\boldsymbol{r}_{\text{opt}}$ at the 25% mark in the vessel, $r(\lceil n/4 \rceil)$.

(b) The last radius value in $\boldsymbol{r}_{\text{opt}}$ is determined by considering the last change point, $\psi_m$, or the last radius observation, $r(n)$. Which option is used depends on the data structure. We compare the slopes of the second-to-last, $slp_m$, and last, $slp_{m+1}$ segments as these slopes differ significantly for many vessels, particularly in small vessels with diameters close to the image resolution (Cui et al., 2019; Lesage et al., 2009). To compare these slopes, we compute

$$s_a = \left| \frac{slp_{m+1} - slp_m}{slp_m} \right|. \tag{14}$$

If $s_a \leq \xi$, we let $r_{\text{opt}}(k) = r(\psi_m)$ and otherwise, $r_{\text{opt}}(k) = r(m)$.

*<u>Case 3</u>:* For vessels with no change points (typically short vessels), we let $\boldsymbol{r}_{\text{opt}} = r(1, n)$.

***Tapering vessels.*** The methodology for determining $\boldsymbol{r}_{\text{opt}}$ for tapering vessels is similar. Instead of examining the segment with the minimum slope, we examine the longest segment $l_{\max} = \max\left(\text{length}(sec_1, sec_2, \dots, sec_{\text{m+1}})\right)$. Identification of the optimal radius segment for tapering vessels depends on four parameters: the threshold for fitting $\boldsymbol{r}_{\text{opt}}$ from a single segment (here $0.4\,n = 40\%$ of the total vessel length), the parameters used in case 2 for non-tapering vessels ($n/4 = 25\%$ and $n/2 = 50\%$), and the threshold parameter $\xi$ (for this study, $\xi = 2.8$).

*<u>Case 1</u>:* If the longest segment includes more than 40% of the total vessel length, we let $\boldsymbol{r}_{\text{opt}} = l_{\max}$. The threshold $0.4\,n = 40\%$ is one of the four input parameters.

*Case 2:* For vessels not included in case 1, we determine the first and last points in $\boldsymbol{r}_{\text{opt}}$ as described in case 2 for non-tapered vessels.

*Case 3:* For vessels with no change points (typically short vessels), we let $\boldsymbol{r}_{\text{opt}} = r(1, n)$. Note these vessels are assigned a non-tapering radius.

*Case 4:* Vessels with optimal radius segment not identified in cases 1—3 , an exponential is fit of the same form described above (Equation (12)). The optimal inlet radius is determined from $r(1)$, and the optimal outlet radius is determined from $r(n)$.

**Step 2: Network Consistency Check.** For healthy subjects, vascular networks satisfy that the radius of the parent vessel is larger than the radii of the daughter vessels $r_p > r_{\boldsymbol{d}}$. In most cases, a parent splits into two daughter vessels and very rarely (less than 2%) splits into three (Chambers, 2022). In addition, the combined area of the daughters is greater than the area of the parent vessel, such that $\sum A_{\boldsymbol{d}} > A_p$ (Zamir, 1978; Murray, 1926; Uylings, 1977; Townsley, 2012). In vessels where these conditions are not satisfied, we adjust the radii depending on the vessel's position within the network. If one terminal daughter, $d_1$, violates the assumption, we let $r_{d_1} = \alpha\, r_p$ and if two terminal daughters violate the assumption, we let $r_{d_1} = \alpha\, r_p$ and $r_{d_2} = \beta\, r_p$. Internal vessels or terminal junctions with trifurcations violating these conditions are manually assigned new radius values from inspection of the data.

**Algorithm 1:** Given radius data $\boldsymbol{r}, m \leq 3$ change points $\boldsymbol{\psi}$, segments between change points $\boldsymbol{sec}$, and slopes of lines fitted within each segment $\boldsymbol{slp}$, this function identifies the region that best determines the vessel radius, $\boldsymbol{r}_{\text{opt}}$, in straight and tapering vessels.

**Function** *get_ropt*:
  **input:** double: $\boldsymbol{r} = \{r_1, \dots, r_n\}, \boldsymbol{\psi} = \{\psi_1, \dots, \psi_m\}, \boldsymbol{slp} = \{slp_1, \dots, slp_{m+1}\}, \boldsymbol{sec} =$
  $\{sec_1, \dots sec_{m+1}\}$; boolean: taper
  **output:** $\boldsymbol{r}_{\text{opt}}$
  **if** *taper == false* **then**
    **if** $m = 0$ **then**
      $\boldsymbol{r}_{\text{opt}} = \boldsymbol{r}$
    **else**
      $\boldsymbol{slp}_{\text{min}} = \min(\{|slp_2|, \dots, |slp_{m+1}|\})$
      $n_s = \text{length}(slp_{min}), k = \psi(slp_{min} - 1)$
      **if** $n_s/n \geq 0.25$ **then**
        $\boldsymbol{r}_{\text{opt}} = r(k, k + n_s)$
      **else if** $\psi_1/n \geq 0.25$ then
        $\boldsymbol{r}_{\text{opt}} = get_slopes(r, slp, \psi_1, \psi_m)$
      **else**
        $\boldsymbol{r}_{\text{opt}} = get_slopes(r, slp, \psi_2, \psi_m)$
      **end**
    **end**
  **else if** taper == true **then**
    $l_{max} = \max(\text{length}(sec))$ $, k = \psi(l_{\text{max}} - 1)$
    **if** $l_max/n \geq 0.4$ **then**
      $\boldsymbol{r}_{\text{opt}} = r(k{:}\, k + l_{max})$
    **else if** $\psi_1/n \geq 0.4$ **then**
      $\boldsymbol{r}_{\text{opt}} = slopes(r, slp, \psi_1, \psi_m)$
    **else if** $\psi_2/n \geq 0.5$ **then**
      $\boldsymbol{r}_{\text{opt}} = get_slopes(r, slp, \lceil n/4 \rceil, \psi_m)$
    **else**
      $\boldsymbol{r}_{\text{opt}} = get_slopes(r, slp, \psi_2, \psi_m)$
    **end**
  **end**

**Algorithm 2:** Given radius data $\boldsymbol{r}$, slopes of linear fits between change points $\boldsymbol{slp}$, indices $i_1, i_2$, and a threshold, this algorithm returns the optimal radius section.

**Function** *get_slopes*:
  **input:** double: $\boldsymbol{r} = \{r_1, \ldots, r_n\}, \boldsymbol{slp} = \{slp_1, \ldots, slp_{m+1}\}, \xi = 2.8$
  integer: $i_1, i_2$
  **output:** $\boldsymbol{r}_{\text{opt}}$
  **if** $|(slp_{m+1} - slp_m)/slp_m| \geq \xi$ **then**
    $\boldsymbol{r}_{\text{opt}} = \boldsymbol{r}(i_1, i_2)$
  **else**
    $\boldsymbol{r}_{\text{opt}} = \boldsymbol{r}(i_1 : n)$
  **end**

## Fluid Dynamics Model

We predict hemodynamics in each vessel using a 1D fluid dynamics model that relates blood pressure, flow, and cross-sectional area. The model is separated into two domains predicting hemodynamics in (1) large arteries with geometry determined from images and (2) small arteries represented by structured trees. Similar to previous studies (Taylor-LaPole et al., 2023; Bartolo et al., 2022; Olufsen et al., 2000; Colebank et al., 2019), in the large vessels, we solve a non-linear unsteady 1D approximation of the Navier–Stokes equations, enforcing conservation of mass and momentum, and in the small vessels, we solve a linearized wave equation model. Parameters for each domain are listed in Table 1 and the supplement.

Table 1: Parameters used in the fluid dynamics model for the pulmonary and aortic networks.

| Quantity | Pulmonary | Aorta |
|---|---|---|
| $T$ (s) | 0.85 | 0.8 |
| $\rho$ (g/cm$^3$) | 1.055 | 1.055 |
| $\mu_L$ (g/cm/s) | 0.032 | 0.032 |
| $\mu_S(r_0)$ | Equation (19) | Equation (19) |
| $k_1^L$ (g/cm/s$^2$) | $2.5 \times 10^6$ | $5.0 \times 10^6$ |
| $k_2^L$ (1/cm) | −15 | −25 |
| $k_1^L$ (g/cm/s$^2$) | $6.4 \times 10^4$ | $6.0 \times 10^5$ |
| $k_1^S$ (g/cm/s$^2$) | $2.5 \times 10^7$ | $5.0 \times 10^6$ |
| $k_2^S$ (1/cm) | −15 | −20 |
| $k_3^L$ (g/cm/s$^2$) | $8.0 \times 10^5$ | $1.0 \times 10^5$ |
| $r_{\min}$ (cm) | 0.001 | 0.001 |
| $(\alpha, \beta)$ | (0.88,0.697) | (0.90,0.60) |
| $\ell rr$ | 15.75 | 50 |

**Large Vessels.** Blood pressure $p(x,t)$ (mmHg), flow $q(x,t)$ (cm$^3$/s), and area deformation $A(x,t)$ (cm$^2$) are predicted under the assumptions that each vessel can be represented by a deformable axisymmetric cylinder with an impermeable wall. We assume that blood is incompressible, Newtonian, viscous, and homogeneous and that the flow is irrotational and laminar. Under these assumptions, the conservation of mass and balance of momentum are given by

$$\frac{\partial A}{\partial t} + \frac{\partial q}{\partial x} = 0$$
$$\frac{\partial q}{\partial t} + \frac{\partial}{\partial x}\left(\frac{q^2}{A}\right) + \frac{A}{\rho}\frac{\partial p}{\partial x} = -\frac{2\pi \nu R}{\delta}\frac{q}{A} \tag{15}$$

where $\mu_L$ is viscosity, $\nu = \mu_L/\rho$ is the kinematic viscosity, $R$ (cm) is the radius, $t$ (s) denotes the temporal coordinate, and $x$ (cm) is the axial position. We assume a flat velocity profile with a linearly decreasing boundary layer (Nichols et al., 2011; Pedersen et al., 1993; Pries et al., 1992; Olufsen et al., 2000) with thickness $\delta = \sqrt{\nu T/2\pi}$ (cm), where $T$ (s) is the length of the cardiac cycle, i.e.,

$$u_x(r,x,t) = \begin{cases} \bar{u}_x, & r < R - \delta \\ \dfrac{\bar{u}_x}{\delta}(R - r), & R - \delta < r \leq R. \end{cases} \tag{15}$$

To close the system of equations, we assume that the vessel wall can be modeled as linearly elastic and isotropic (Qureshi et al., 2019) under which pressure and area can be related as

$$p(x,t) - p_0 = \frac{4}{3}\frac{Eh}{r_0}\left(\sqrt{\frac{A_0}{A}} - 1\right), \qquad \frac{Eh}{r_0} = k_1^L \exp(k_2^L r_0) + k_3^L \tag{16}$$

where $E$ (g/cm/s$^2$) is Young's modulus, $h$ (cm) is the vessel wall thickness, $p_0$ (mmHg) is a reference pressure, $A_0$ (cm$^2$) is the cross-sectional area, and $R_0$ (cm) the radius of the vessel determined by our algorithm. To account for changes in wall composition, the vascular stiffness is modeled as exponentially increasing with vessel radius. In the second equation $k_1^L$ (g/cm/s$^2$), $k_2^L$ (1/cm), and $k_3^L$ (g/cm/s$^2$) are constants based on literature (Colebank et al., 2021a; Paun et al., 2020; Taylor-LaPole et al., 2023).

At the inlet of the root vessel, we prescribe an inflow waveform extracted from data. At the MPA, we specify a flow waveform using data from SimVascular (Updegrove et al., 2017). In the aortic vasculature, a flow waveform at the root of the ascending aorta is extracted from 4D-MRI data provided by Baylor College of Medicine and Texas Children's Hospital. At junctions, we enforce continuity of pressure and conservation of flow between the parent vessel $p$ and the $k$ daughters, $\mathbf{d} = (d_1, d_2, \ldots, d_\#)$)

$$p_p(L,t) = p_{d_l}(0,t), \qquad i = 1,2,\ldots,l, \qquad \text{and} \quad q_{p(L,t)} = \sum_{l=1}^{k} q_{d_l}. \tag{17}$$

**Small Vessels.** In small vessels with radii smaller than image resolution, viscous forces are dominant, allowing us to neglect the nonlinear inertial terms and linearize Equations (15), as done in previous studies (Olufsen et al., 2000). In the frequency domain, the conservation of mass and momentum equations become

$$i\omega Q + \frac{A_0\left(1 - F_J\right)}{\rho}\frac{\partial P}{\partial x} = 0, \qquad F_J = \frac{2J_1(w_0)}{w_0 J_0(w_0)}$$
$$i\omega CP + \frac{\partial Q}{\partial x} = 0, \qquad C = \frac{\partial p}{\partial A} \approx \frac{3}{2}\frac{r_0}{Eh}, \tag{18}$$

where $J_0, J_1$ are the zeroth and first order Bessel functions, $w_0^2 = i^3 r_0 \omega/\nu$, $C$ is vessel compliance, and $Eh/r_0$ has the same form as Equation (16) but with parameters $k_1^S, k_2^S$, and $k_3^S$. Viscosity becomes more significant for smaller vessels (Pries et al., 1992), so in this domain, we compute viscosity as a function of vessel radius as

$$\mu_S(r_0) = \frac{D\mu_L}{3.2}\left(1 + D\left(6e_0^{-0.17r} + 3.2 - 2.44e_0^{-0.12r_0^{0.645}} - 1\right)\right), \tag{19}$$

where $D = (2r_0/(2r_0 - 1.1))^2$ is the relative viscosity at a hematocrit level of 0.45. Solutions of these equations allow us to compute $Z(0,\omega) = P(0,\omega)/Q(0,\omega)$ at the beginning of each vessel as a function of the impedance at the end of the vessel as

$$Z(0,\omega) = \frac{ig_\omega^{-1}\sin(\omega L/c) + Z(L,\omega)\cos(\omega L/c)}{\cos(\omega L/c) + ig_\omega Z(L,\omega)\sin\left(\omega L/c\right)}. \tag{20}$$

Using same junction conditions as the large vessels, the impedance at the root of the structured tree is computed recursively. The combined system of equations is solved numerically using the two-step Lax-Wendroff finite difference scheme (Colebank et al., 2019, 2021a; Bartolo et al., 2022; Taylor-LaPole et al., 2023; Olufsen et al., 2000; Chambers et al., 2020).

## Statistical Analysis and Uncertainty Quantification

To determine uncertainty associated with image segmentation, we perform a one-way analysis of variance (ANOVA), comparing five segmentations of the pulmonary vasculature and ten segmentations of aortic vasculature prior to adjusting the labeled trees (i.e, before modifying junctions and extracting radii based on change points). This statistical test determines if significant differences exist among the variance of different samples (Kauffman and Schering, 2007). We conduct the ANOVA tests in a vessel-specific manner to determine if there are statistically significant differences in vessel radii and length from the different 3D renderings and centerlines of the same subject. The ANOVA calculations are performed in R using the function anova() from the R/stats package (R Core Team, 2013) (Table 2). More details about ANOVA can be found in the supplemental material (S4).

To examine variability within the radius data, we calculate coefficients of variation as $CV = \sigma_{\hat{r}}/\hat{r}$ for vessel radii in each segmentation (Grafton et al., 2022). Here, $\hat{r}$ is the mean of the optimal radius region and $\sigma_{\hat{r}}$ is its standard deviation.

We also compare kernel density estimations (KDE) of vessel radii for each segmentation. The kernel density represents the data using the radii's probability density function (PDF). These KDEs are plotted to visualize the estimated distribution of the radii throughout the segmented vasculatures (Weglarczyk, 2018). We use MATLAB's ksdensity() to convert the radii from each segmentation into a PDF (MathWorks, 2022).

## Simulations

To assess uncertainty in image segmentation and 1D network generation, we perform two fluid dynamic simulation types. First, we compare multiple networks generated from the same image, assessing the impact of segmentation uncertainty and pruning. Then we study the impact of variation in junction placement and vessel radius within a representative network.

**3D Rendering Variation.** The pulmonary network was segmented five times and the aortic network was segmented ten times by the same user. For both, we used the same threshold parameters and smoothing type. All aortic networks have the same number of vessels and connectivity, but these features differ in the pulmonary trees. Thus, we examine the effect of discrepancy in vessel count and the impact of pruning.

**Variations in Junction and Radius Estimations.** A representative network is selected to study the influence of junction placement and vessel radius on hemodynamics. We demonstrate the impact of altering the location of the junction node, which changes the length and radii considered for each vessel in the system. Once junction nodes are adjusted, we explore how radius uncertainty propagates to fluid dynamics. To do so, we fit a normal distribution $N(\hat{r}, \sigma_{\hat{r}}^2)$ to the optimal region representing the vessel radius (Figure 6). Sampling from a normal distribution is justified by systematically analyzing PDF in the optimal region. In vessels with reassigned radii, we sample from the distribution $N(k\hat{r}, k\sigma_{\hat{r}}^2)$ , where $\kappa = \hat{r}_d/\hat{r}_p$ is a parent-daughter scaling factor. Within the normal distribution that represents possible radii values, we sample 1000 times in the chosen representative network and 100 times in other segmentations.

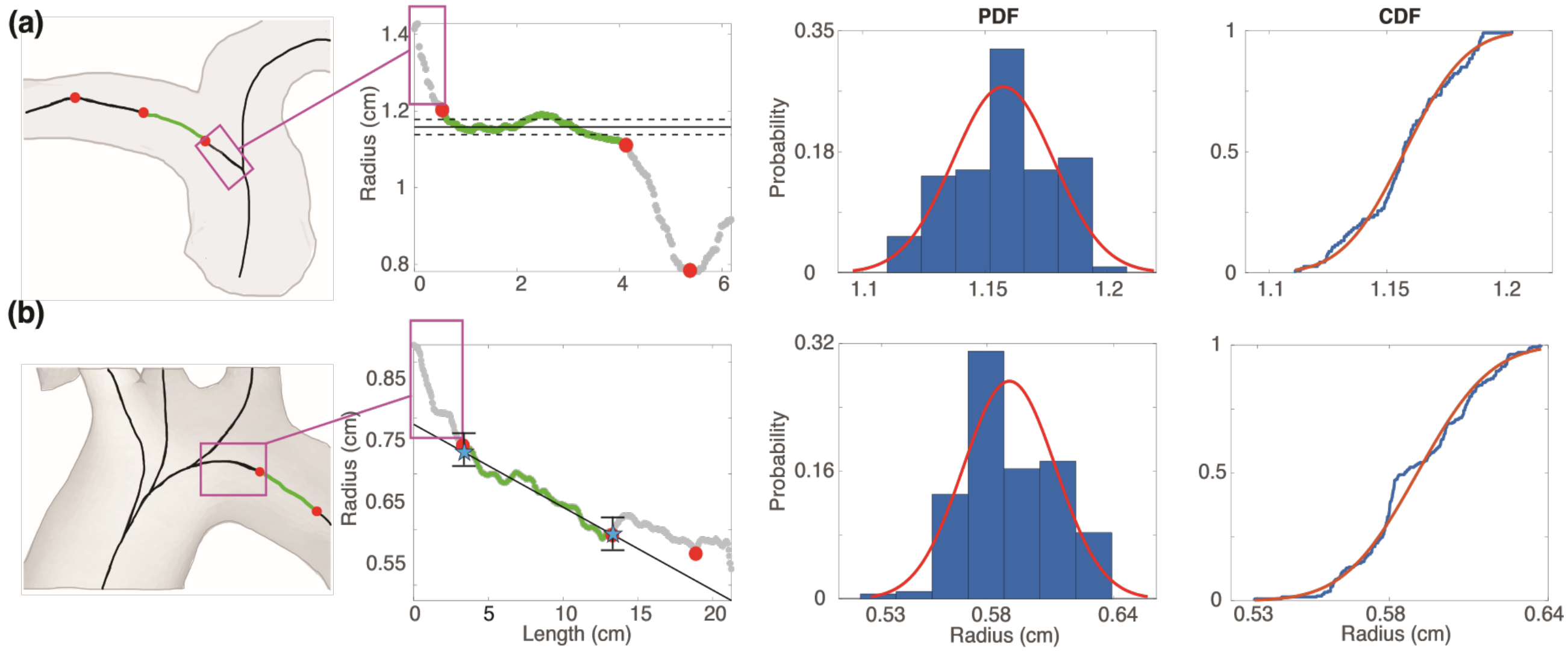

Figure 6: Radius extraction for (a) pulmonary and (b) aortic arteries. Change points (red circles) denote locations where the vessel radius changes significantly. The segment used to determine the vessel radius (green) avoids the junction region (pink box) and rapidly changing radii at the end of the vessel. In the aorta, estimated inlet and outlet

radii are denoted by blue stars. The probability (PDF) and cumulative (CDF) density functions for the standard deviation of optimal radii (green section) are shown in the right panels.

# RESULTS

**Variation in 3D Rendering.** One-way ANOVA (Table 2) of the radius and length obtained from the raw networks generated from VMTK (before junction and radius adjustment) show that the variability between the dimensions generated from each segmentation is more significant than what would be expected by solely random chance (Kauffman and Schering, 2007). The ANOVA analysis on seven pulmonary arteries matched in each segmentation shows statistically significant differences in length and radii. In the aortic network, the radius of the innominate, brachial, right common carotid (RCC), and right subclavian arteries differ significantly. The length of the left common carotid (LCC), brachial, left vertebral, and RCC also varies significantly between segmentations. This analysis demonstrates that the raw data obtained from VMTK is uncertain despite having the same user segment all images using standard segmentation parameters, methods, and software. This justifies the need to generate a method to extract length and radii values for vessels in a network while accounting for uncertainty.

Table 2: One-way ANOVA test results predict the error sum of squares (SSE) and p-values for seven vessels in the pulmonary vasculature and eight vessels from the aortic network that can be matched in each segmentation. For continuity, all aortic vessels (ascending, arch I, arch II, and descending) were treated as one exponentially tapering vessel. P-values in bold text are statistically significant under a 0.05 threshold.

| | **Pulmonary** | | | **Aorta** | | |
|---|---|---|---|---|---|---|
| | **Name** | **SSE** | **p-value** | **Name** | **SSE** | **p-value** |
| Radius ANOVA | MPA | 2.82 | < **0.001** | Aorta | 0.94 | 0.11 |
| | RPA | 5.45 | < **0.001** | Inominate | 2.73 | < **0.001** |
| | LPA | 3.55 | < **0.001** | LCC | 0.14 | 0.97 |
| | RIA | 3.20 | < **0.001** | L Subclavian | 0.16 | 0.32 |
| | LIA | 13.7 | < **0.001** | L Brachial | 0.10 | **0.0093** |
| | RTA | 1.05 | < **0.001** | L Vertebral | 0.004 | 0.10 |
| | LTA | 18.6 | < **0.001** | RCC | 0.12 | < **0.001** |
| | | | | R Subclavian | 0.12 | < **0.001** |
| Length ANOVA Results | MPA | 246 | < **0.001** | Aorta | 652 | 0.*40* |
| | RPA | 178 | < **0.001** | Innominate | 0.77 | 0.99 |
| | LPA | 74.0 | < **0.001** | LCC | 298 | **0.02** |
| | RIA | 37.8 | < **0.001** | L Subclavian | 4.00 | 0.99 |
| | LIA | 195 | < **0.001** | L Brachial | 43.1 | **0.047** |
| | RTA | 14.3 | < **0.001** | L Vertebral | 58.6 | **< 0.001** |
| | LTA | 32.0 | < **0.001** | RCC | 371 | **< 0.001** |
| | | | | R Subclavian | 9.37 | 0.83 |

Main (MPA), right (RPA) and left (LPA) pulmonary arteries; Right (RIA) and left (LIA) interlobular arteries; and right (RTA) and left (LTA) trunk arteries.

**Network Pruning.** Figure 7 shows the impact of standardizing the network size between segmentations. Results show predictions of pressure and flow at the midpoint of the MPA, RPA, and LPA. Before pruning, predicted pressures vary significantly between segmentations with systolic differences of over 15 mmHg (Figure 7, solid lines). The segmentation with the highest number of vessels pre-pruning has a pressure range between 18 − 31 mmHg. The segmentation with the smallest number of vessels has a pressure range of 5 − 15 mmHg. Once all vessels are pruned to 167 vessels, the pressure range between each segmentation lessened to about 5 mmHg (Figure 7, dashed lines). Since a fixed flow profile is imposed at the inlet to the network, there is significantly less variation in flow than pressure. However, pruning still impacts flow predictions favorably. Before pruning, the peak RPA flow varies by approximately 100 mL/s and after pruning, the variation becomes less than 50 mL/s. Flow discrepancies in the MPA and LPA flow are smaller. These results are summarized in Table 3.

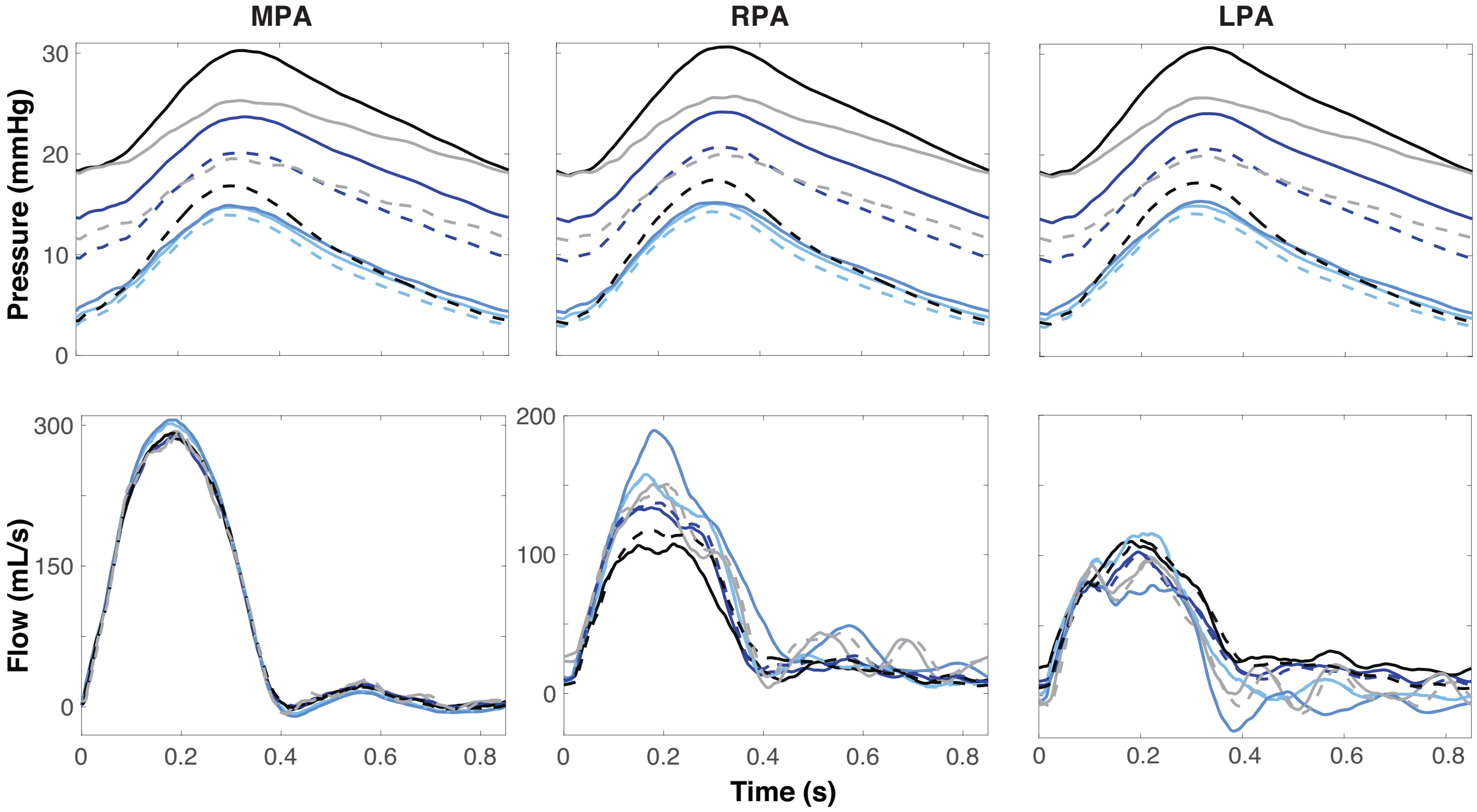

Figure 7: Pressure and flow dynamics over a cardiac cycle for each segmentation at the midpoint of the MPA, RPA, and LPA in the pulmonary arteries before (solid lines) and after (dashed lines) standardizing network size through pruning. Each color represents a distinct segmentation.

**Variance in VMTK Radius Estimates.** Figure 8 shows the CV and KDE for the pulmonary and aortic vasculatures. The largest vessels in the pulmonary vasculature, with radii between 1−1.5 cm, have relatively small CVs, indicating that the centerlines generated from the segmentation and radius extraction for these vessels have less noise. For vessels with radii less than 0.5 cm, CVs range from 0−0.6, with one outlier reaching a CV of over 1. This vessel has a standard deviation greater than its mean, which can occur when terminal vessels are close to image resolution limits. Smaller vessels consistently have a higher and broader range of CV values in each segmentation.

In the aortic vasculature, CVs range from 0 − 0.32. Vessels with radii less than 0.4 cm and greater than 1 cm have a CV of less than 0.18. Vessels with radii between 0.4−1 cm have CV values greater than 0.27. Similar to the pulmonary arteries, the largest aortic vessels have the smallest CV of less than 0.1. The pulmonary KDE plots have a unimodal, right-skewed distribution with the majority of vessel radii from all segmentations being less than 0.5 cm, peaking at 0.1 cm. Given the variations in segmentations, the radii distribution varies across networks even after pruning. For example, in one segmentation, a radius of 0.1 cm is observed in 17.5% of vessels, while in another, this radii value is found in 12.5%. However, it should be noted that prior to pruning this difference more pronounced with up to 20% variation between networks (Figure 3). However, it should be noted that prior to pruning this difference more pronounced with up to 20% variation between networks (Figure 3). For the aortic networks, the KDE follows a right-skewed bimodal distribution with a major and minor modes at 0.25 and 0.6 cm, respectively. Most aortic vessels have a radius of less than 0.5 cm, with another large portion having a radius between 0.5 − 0.75 cm. Similar to the pulmonary vessels, distinct networks have different radii distributions. Despite that, the radii distribution between segmentations is less than 5%.

Table 3: Range, minimum, and maximum pressure, and flow before and after pruning.

| | | **Pressure (mmHg)** | | | **Flow (mL/s)** | | |
|---|---|---|---|---|---|---|---|
| ***Pulmonary*** | | **Range** | **Min** | **Max** | **Range** | **Min** | **Max** |
| MPA | *Before* | 25 | 5 | 30 | 330 | -15 | 315 |
| | *After* | 20 | 5 | 25 | 320 | -10 | 310 |
| RPA | *Before* | 25 | 5 | 30 | 170 | 10 | 180 |
| | *After* | 20 | 5 | 25 | 150 | 10 | 160 |
| LPA | *Before* | 25 | 5 | 30 | 150 | -30 | 120 |
| | *After* | 20 | 5 | 25 | 130 | -10 | 120 |

MPA: main pulmonary artery, RPA: right pulmonary artery, LPA: left pulmonary artery.

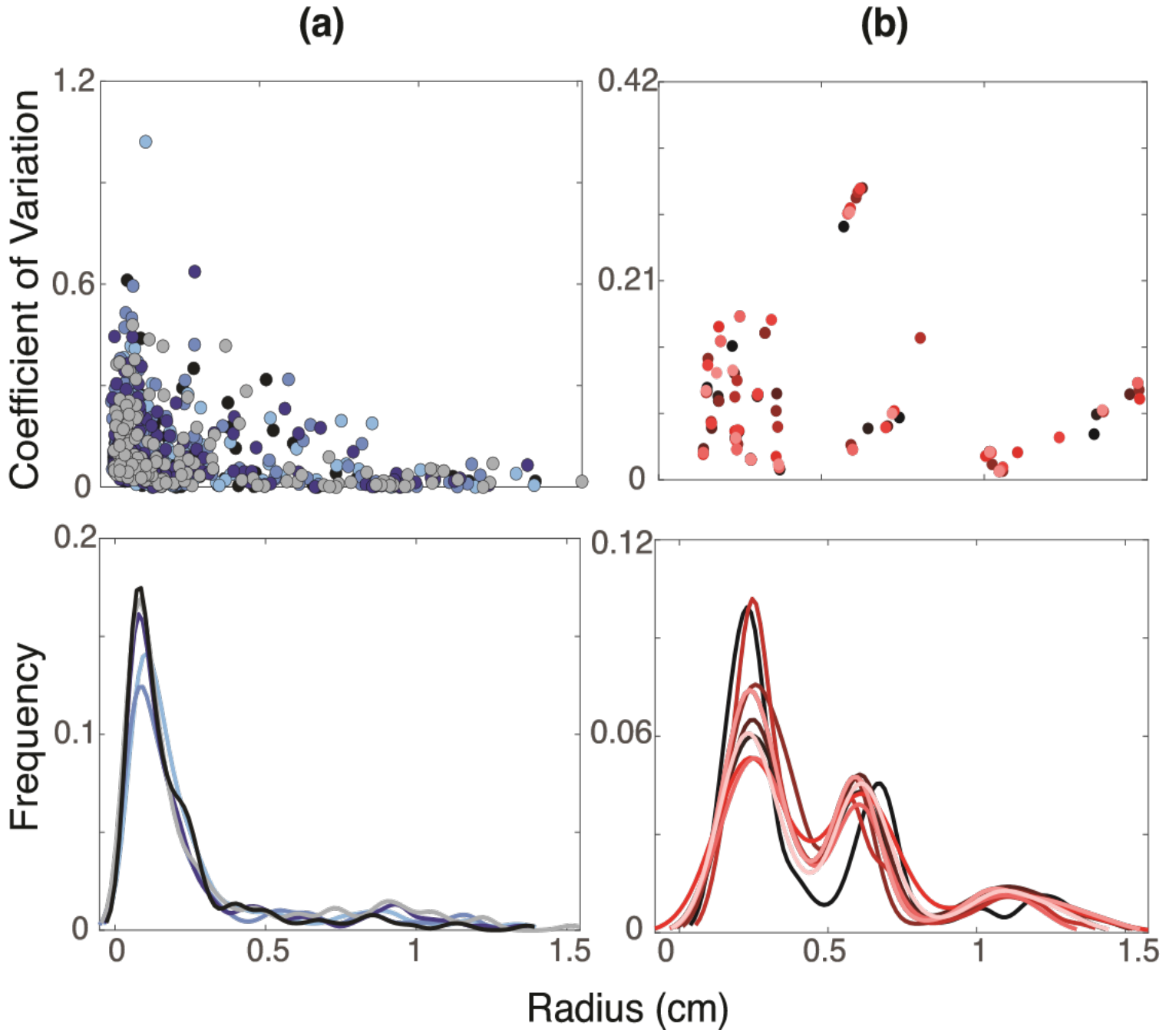

Figure 8: Coefficient of variation and kernel density estimates for (a) pulmonary arteries and (b) aortic vessels. Each color represents a distinct segmentation.

**Junction Node Variation.** Pressure and flow waveforms calculated at vessel midpoints before (solid lines) and after (dashed lines) modifying junctions are shown in Figure 9. The variation before and after modifying junctions is significantly smaller in the pulmonary vessels compared to the aorta. The pressure varies by 1 − 3 mmHg in the MPA, RPA, and LPA. The difference between flow predictions is also negligible in the pulmonary compared to the aortic vasculature, but there is a distinct effect on the shape of the flow predictions. Before modifying junctions, pulse pressure in the aortic vasculature is about 80 mmHg, ranging from 40 − 120 mmHg. After this change, the pulse pressure is 50 mmHg, ranging from 50 − 110 mmHg. Before adjusting the junctions, the flow to the LCC and brachial arteries are higher, whereas, after this modification, more flow is directed to the descending aorta. Table 4 lists the range, minimum, and maximum pressure, and flow, before and after adjusting the junctions. Wave reflections are also impacted by modifying junction nodes, particularly in the LCC. This is likely a result of reduced flow to the LCC after the junction modification. This difference is amplified in the aortic segmentation reported with a solid black line with a sizeable negative dip and one large reflection around 0.6 s. After modifying junctions, we observe one steady reflection between 0.5 − 0.7 s. The opposite trend is apparent in the brachial artery, where predictions before junction modification show one reflection and those after show two minor reflections.

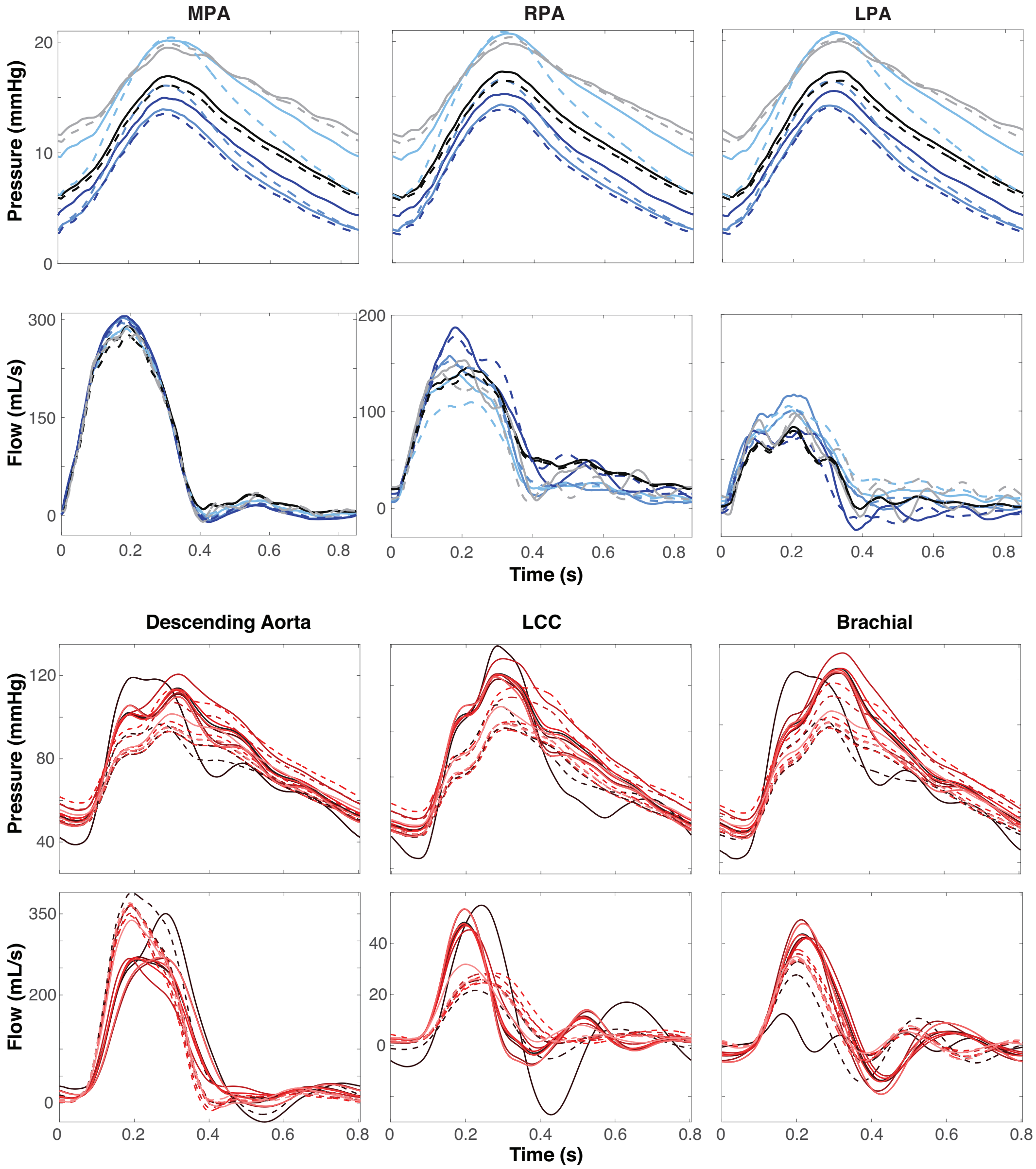

Figure 9: Pressure and flow dynamics over a cardiac cycle for each segmentation at the midpoint of (a) the MPA, RPA, and LPA in the pulmonary arteries and (b) the descending aorta, LCC, and brachial artery before (solid lines) and after (dashed lines) modifying VMTK-derived junctions. Each color represents a distinct segmentation.

Table 4: Range, minimum, and maximum, pressure, and flow, before and after junction modification.

| | | **Pressure (mmHg)** | | | **Flow (mL/s)** | | |
|---|---|---|---|---|---|---|---|
| ***Pulmonary*** | | **Range** | **Min** | **Max** | **Range** | **Min** | **Max** |
| MPA | *Before* | 15 | 5 | 20 | 320 | -10 | 310 |
| | *After* | 15 | 5 | 20 | 310 | -10 | 300 |
| RPA | *Before* | 15 | 5 | 20 | 160 | 10 | 170 |
| | *After* | 15 | 5 | 20 | 165 | 10 | 175 |
| LPA | *Before* | 15 | 5 | 20 | 200 | -50 | 150 |
| | *After* | 15 | 5 | 20 | 80 | -10 | 160 |
| ***Aortic*** | | | | | | | |
| Desc Aorta | *Before* | 80 | 40 | 120 | 400 | -50 | 350 |
| | *After* | 60 | 50 | 110 | 275 | -25 | 250 |
| LCC | *Before* | 90 | 35 | 125 | 75 | -25 | 50 |
| | *After* | 60 | 50 | 110 | 30 | 0 | 30 |
| Brachial | *Before* | 85 | 35 | 120 | 70 | -20 | 50 |
| | *After* | 60 | 50 | 110 | 40 | -10 | 30 |

MPA: main pulmonary artery, RPA: right pulmonary artery, LPA: left pulmonary artery, Desc Aorta: descending aorta, LCC: left common carotid, brachial: brachial artery.

**Change Point Analysis.** Figures 5 and 6 depict the results of the change point analysis used to identify vessel radii. We choose a maximum of three change points in each vessel to balance model complexity and interpretability. We analyze nearly 1000 vessels within five segmentations of pulmonary arteries (167 vessels each) and ten of the aortas (13 vessels each). We find that 9% of pulmonary arteries and 33% of aortic vessels have either zero, one, or two change points. The rest of the vessels have three change points. After combining change point detection with our radius extraction protocol, less than 1% of vessels require post-processing during the network consistency check. This demonstrates the robust nature of the framework we have developed.

**Radius Variation.** Figure 10 shows how varying vessel radii impacts pressure and flow waveforms in a representative network. Pulmonary pressure predictions in the MPA, RPA, and LPA range from 10−25 mmHg in systole and 0−15 mmHg in diastole. In the MPA, there is a slight variation in flow predictions in systole ranging from approximately 275 − 290 mL/s. In the RPA and LPA, variations are more pronounced, ranging from 100 − 175 mL/s and 50 − 100 mL/s, respectively. Aortic pressures range from 90 − 160 mmHg in systole and 50 − 100 mmHg in diastole. Flow predictions in the descending aorta range from 250−350 mL/s in systole, while the LCC and brachial range from 30 − 40 mL/s. Table 5 lists the range, maximum, and minimum pressure, and flow predicted from varying radius .

## DISCUSSION

This study devises novel algorithms for generating labeled directed trees from medical images. The algorithms are demonstrated on trees extracted from CT and MRA images of the pulmonary arteries and aortic vasculature, respectively. To study variation to image segmentation, we extract multiple trees for each vasculature from the same image. We also develop a robust method for accurately placing nodes at vessel junctions and use change points to determine segments within each vessel that represent its true radius. From this segment, we determine the vessel radius and its standard deviation. We examine hemodynamics in generated trees, exploring the impact of variation over segmentations, radii, and length of each vessel. Our results show that by pruning vascular networks and carefully extracting vessel radii and length using the algorithms designed in this study, users can significantly reduce fluctuations in hemodynamic predictions, improving the methodology for patient-specific modeling.

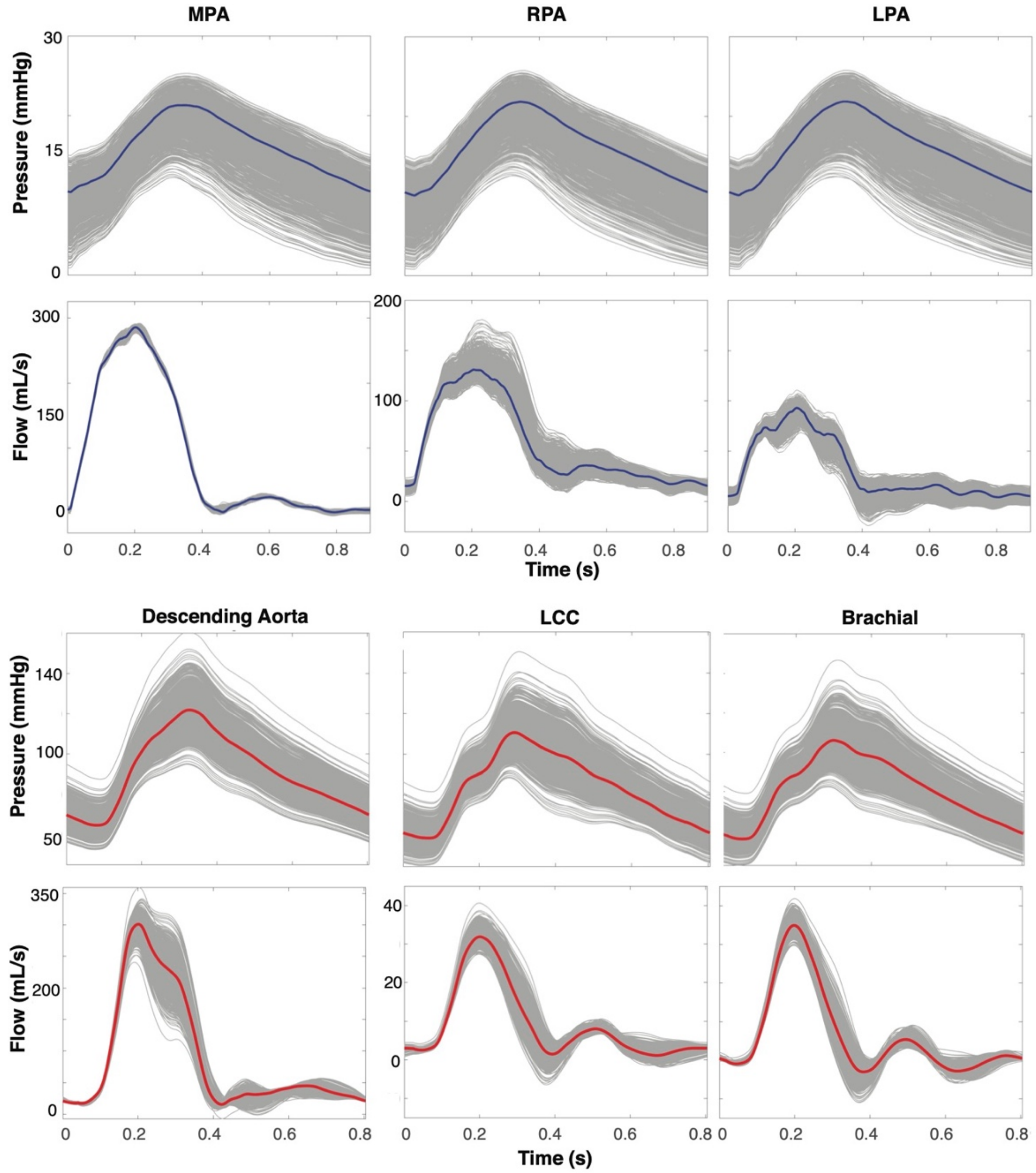

Figure 10: Pressure and flow dynamics over a cardiac cycle at the midpoint of the (a) MPA, RPA, and LPA in the pulmonary arteries and (b) descending aorta, LCC, and brachial artery. Gray lines (1000) denote simulations with radii sampled from a normal distribution based on the mean and standard deviation of each vessel's extracted radius, and the colored line (1) represents the simulations with radii set at the mean radius based on our extraction procedure. These simulations were based on a representative segmentation of the pulmonary and aortic vasculatures.

**Analysis of Raw Networks from VMTK.** Analysis of the raw networks generated by VMTK from the 3D rendered surfaces shows that the CV is higher in the pulmonary vasculature compared to the aortic, especially in the smallest vessels (Figure 8). This is an expected finding since the aortic network only consists of relatively large vessels, while the pulmonary network has many small vessels near image resolution. Moreover, high deviation is expected since pulmonary tree segmentation requires manual components.

The bimodal nature of the KDE plots in aortic vessels (Figure 8) stems from the change in the size of the vessels analyzed. The ascending aorta, arch I, arch II, and descending aorta have significantly larger radii than the branching head and neck vessels. In contrast, the pulmonary vasculature has a rapidly branching structure, giving rise to a unimodal right-skewed KDE. These findings reflect uncertainty in the radii extracted from centerlines, obtained from the segmentations of MRI and CT images. The importance of quantifying uncertainty associated with quantities extracted from image segmentation through the use of KDEs and PDFs has been supported by several studies. Baumgartner et al. (2019) generate PDFs for each segmentation and compare it to the distribution of the actual image. They argue that knowing these PDFs can greatly improve image segmentation accuracy (Baumgartner et al., 2019). KDEs have also been used for machine-learning applications of image segmentation. Wang et al. (2013) leverage KDEs of abdominal blood vessels from CT angiograms to learn target distributions and estimate vessel locations in segmentations of the same vessels in different patients. The present study uses KDEs to understand vessel distributions within and between segmentations, and we note that this analysis can be a powerful tool for gaining information about vessel uncertainty.

It is well known that limited image resolution, segmentation inaccuracies, or inexactness in centerline generation (Lins et al., 2020; Sharma and Aggarwal, 2010; van Rikxoort and van Ginneken, 2013; Buelow et al., 2005) impact 3D rendering. Other factors, including noise, poor contrast, and blurred boundaries, also increase the difficulty of extracting the smallest vessels (Cui et al., 2019; Lesage et al., 2009; Hirsch et al., 2012). In the study by Colebank et al. (2019) examining image segmentation uncertainty, the uncertanty increases as vessel radius decreases, suggesting that small vessels are more sensitive to segmentation parameters. Van Horssen et al. (2016) postulate that small vessels are more likely underestimated due to image resolution. Schwarz et al. (2020) found that in the human coronary arterial circulation, the smallest vessels are subject to segmentation errors, resulting in triangular loops and spurious side branches that must be corrected in post-processing. These studies validate our findings that the aortic networks with larger vessels have less variation than the small vessels in the rapidly branching pulmonary network. This variation is not a flaw of the software used to generate segmentations and networks, but points to the need for post-processing when using image-generated networks for hemodynamic predictions.

**Network Standardization.** Standardizing networks to have a consistent number of vessels is imperative to enable comparison. After pruning pulmonary networks, we find that fluid dynamic differences stemming from segmentation variation become less apparent. An alternative way to account for large vessel network size differences is to adjust structured tree parameters, $\alpha$ and $\beta$, generating networks with similar total vessel count and a similar total vessel area or volume. However, this study investigates changes in geometric quantities, such as radius and length, keeping parameters constant to emphasize the uncertainty caused by vascular geometry. Thus, we prune networks rather than adjusting $\alpha$ and $\beta$ parameter values, though this can be investigated in future studies. Miller et al. (2023) argue that networks with disparate branch numbers are an imaging consequence rather than a biological factor, and thus, pruning is essential when comparing different segmentations. In addition, they find that more information about networks can be gained by applying pruning techniques (Miller et al., 2023). Colebank et al. (2019) find that choices in segmentation parameters greatly impact the number of vessels in the network and that network size affects the model output. Since mean pressure and flow metrics are used to diagnose diseases, such as pulmonary hypertension (Hoeper et al., 2013), standardizing results and assessing uncertainty is critical. This also highlights the importance of determining the minimum number of imaged arteries needed for reliable diagnostic results. If a subset of larger vessels can yield valid, comparable results, excluding smaller vessels that exhibit drastic imaged-related uncertainties may be beneficial. This idea has been studied by several groups, who use surrogate or reduced order models to lessen computational complexity (Paun et al., 2021; Borowska et al., 2023; Ragkousis et al., 2016; Epstein et al., 2015).

Table 5: Range, minimums, and maximums of pressure and flow predictions comparing the "true" geometry to predictions that are from sampled geometries. The "true" geometry comes from the mean of the optimal section selected by change points and the sampled comes from sampling from a normal distribution around the mean.

| | | Pressure (mmHg) | | | Flow (mL/s) | | |
|---|---|---|---|---|---|---|---|
| ***Pulmonary*** | | **Range** | **Min** | **Max** | **Range** | **Min** | **Max** |
| MPA | *True* | 10 | 13 | 23 | 280 | 0 | 280 |
| | *Sampled* | 25 | 0 | 25 | 285 | 0 | 285 |
| RPA | *True* | 10 | 13 | 23 | 110 | 15 | 125 |
| | *Sampled* | 25 | 0 | 25 | 170 | 10 | 180 |
| LPA | *True* | 10 | 13 | 23 | 100 | 0 | 100 |
| | *Sampled* | 25 | 0 | 25 | 110 | -25 | 115 |
| ***Aortic*** | | | | | | | |
| Desc Aorta | *True* | 60 | 60 | 120 | 400 | 0 | 250 |
| | *Sampled* | 75 | 45 | 160 | 275 | -25 | 250 |
| LCC | *True* | 50 | 60 | 110 | 30 | 0 | 30 |
| | *Sampled* | 110 | 40 | 150 | 55 | -15 | 40 |
| Brachial | *True* | 50 | 60 | 110 | 30 | 0 | 30 |
| | *Sampled* | 110 | 40 | 150 | 60 | -20 | 40 |

MPA: main pulmonary artery, RPA: right pulmonary artery, LPA: left pulmonary artery, Desc Aorta: descending aorta, LCC: left common carotid, brachial: brachial artery.

**Junction Node Placement.** The placement of nodes within junctions impacts both the length and radii of vessels. Large vessels, especially the aorta, are known to taper along their length (Caro et al., 1978). Due to the rapid change in vessel radii, junction node placement in these vessels is essential. Our algorithm addresses this by shifting the VMTK-derived nodes to the center of the ostium. As a result, sections of centerline data previously characterized as part of daughter vessels become reclassified as part of the parent. Modifying junction nodes impact hemodynamics in the pulmonary and aortic vasculature (Figure 8) but impacts aortic vessels more significantly. Since aortic vessels are longer and wider than pulmonary vessels, geometric changes resulting from modifying junctions are more pronounced, and this effect is propagated to hemodynamics. In particular, in the descending aorta, one segmentation has a nearly 60 mmHg variation in systole between junction node locations. This is not the case in pulmonary vessels, with an average pressure variation of only 3 mmHg in systole. The wave reflections in the aortic flow predictions are also affected. Abdullateef et al. (2020) investigates the impact of fluctuating vessel radius on wave reflections, finding that as vessel radius decreases, wave reflections increase and begin to overlap. Accurately placing nodes in tapering vessels is crucial, as an incorrect placement can cause wrong reflective wave properties. Due to the well-known interplay between geometry and hemodynamics, generating accurate vessel geometry is imperative. Although this study uses the cylindrical Navier−Stokes equations to model hemodynamics, considering the concept of steady flow within a cylinder, namely Poiseuille's law $q = \pi p r^4 / 8 \mu l$ (Pfitzner, 1976; Demers and Wachs, 2022), allows us to gain insight into how radius and length influence pressure and flow dynamics. Clearly, $q$ and $p$ are directly impacted by $r$ and $l$ in Poiseuille's law. Notably, $r$ is raised to the fourth power, whereas $l$ is raised to the first power. These results are reflected in our study, though it should be noted that changing the junction nodes' location affects the vessel length and radius since nodes with smaller or larger radii may be moved to a different part of the vasculature. The latter, in particular, affects the aortic vasculature, where small and large vessels merge at junctions.

**Change Point Analysis.** We sequentially combine two methods for change point analysis, utilizing R/changepoint to detect the optimal number of change points (Killick and Eckley, 2014) and R/segmented to place them (Muggeo, 2008). Both methods are computationally fast, and their combination enables accurate and automated assessment of vascular data. R/changepoint determines the optimal change point count using mean and variance to detect changes in the data. However, it is an approximation algorithm, only giving an estimated change point placement rather than an exact location (Killick and Eckley, 2014). In contrast, R/segmented accurately places change points through a customizable and flexible algorithm, where users define linear regression models based on prior knowledge of data structure (Muggeo, 2008). This package requires users to input the number of change points detected within each dataset (Muggeo, 2008), so we use the result from R/changepoint as an input. Coupling these two packages allows us to create an automated pipeline for accurate vascular network analysis requiring minimal user input. This is advantageous when comparing multiple vascular datasets, which may encompass thousands of vessels.

By limiting the number of change points to three, we ensure that detected points capture meaningful transitions in a vessel's structure, such as the ostium region or image resolution loss, while preventing the overfitting of the data. It enables consistent and computationally fast analysis, allowing for the development of a robust algorithm for extracting radii. While we could define exactly three change points in all datasets, this leads to inaccurate change point placement in vessels with an optimal count of less than three, as observed in 9% of pulmonary vessels and 33% of aortic vessels.

**Radius Sampling.** Figure 10 shows that vessel radius greatly impacts hemodynamics, following the pattern we expect from Poiseuille's law. Given the predicted pressure and flow values for various radii, a small change in radius causes a large change in hemodynamics. For example, when we perturb the LCC's radius following a distribution N(0.36,0.01), we observe a 70 mmHg range in systolic pressure. This is also shown in a study by Colebank et al. (2019), where even slight changes in vessel dimensions influence pressure and flow. Secomb (2016) notes that small changes in diameter can widely modulate blood flow and emphasizes the need to precisely identify vessel dimensions to understand how blood is distributed throughout the body. We note that pulmonary blood pressures fluctuate throughout the respiratory cycle, evident when using right-heart catheterization for pulmonary artery measurements (Kovacs et al., 2014). Pressure measurements are greatly impacted by patient angle during data collection, changes in altitude, ambulation, and pre-existing heart arrhythmias. Especially when measuring the same subject over time, this can lead to a pressure variation of between 5−15 mmHg (Magder, 2018; Alam et al., 2022). Our pressure range obtained from sampling radii falls within the anticipated range, depending on how much the radius is changed. It is common practice to segment cardiovascular networks to build patient-specific models used as noninvasive methods to monitor and understand disease (Colebank et al., 2021a; Taylor-LaPole et al., 2023; Antiga et al., 2008; Carson et al., 2019; Dobroserdova et al., 2016). Our findings demonstrate that accurate radius measurements are necessary to tune models to patient data accurately, especially when using patient-specific parameters, such as vessel stiffness.

The use of change points to identify the vessel radius and its standard deviation is novel, providing an automated way to identify the vessel radius accurately. Most previous studies either average all nodes within the vessel (Pfaller et al., 2022; Mulder et al., 2011), or assign the vessel radius using a set proportion of the nodes within the vessel without accounting for different vessel structures (Colebank et al., 2021b).

Correct identification of vessel radii is important beyond computational modeling. In the current study, our algorithm is used for identifying vessel radii in vasculature networks informing the geometric domain for computational analysis. However, many studies segment vascular structures from CT and MRI images without using CFD. This is useful when studying diseases and surgical interventions, such as hypoplastic left heart syndrome (HLHS) and chronic thromboembolic pulmonary hypertension (CTEPH). HLHS patients often have enlarged and/or deformed aortas due to reconstructive surgery (Taylor-LaPole et al., 2023). Our algorithm can identify patients with abnormal vessel radii and compare radius distribution to control individuals, such as DORV patients. Another example is quantifying vascular remodeling in patients with CTEPH, including quantifying enlargement of the MPA an acute radius reduction in vessels with ring- or web-like lesions. Computing radii values with uncertainty can provide valuable guidance for surgical interventions, assisting physicians in identifying which vessels with lesions should be targeted.

**Limitations and Future Studies.** This paper outlines a semi-automatic framework to extract and refine patient-specific geometries from medical images. We use a linear regression model to define change points for all vessels. With these change points, we can avoid unclear ostium regions and reliably detect representative radii. However, since we use an exponential function to model tapering radii, we can reduce computational complexity by directly defining change points with an exponential model in those cases (Chen and Gupta, 2011).

In addition, it must be noted that we make several assumptions regarding the evolution of a vessel's radius and the locations of junctions to develop this framework. These assumptions are made based on a thorough analysis of typical vessel structures in large networks and a literature review. However, our framework is highly customizable, so these assumptions can be adjusted to suit the needs of various vascular networks that differ from the aorta and pulmonary arteries, such as the renal, cerebral, and hepatic arteries. Future studies will investigate the impact of specific modeling assumptions on hemodynamic predictions in different systems.

Another limitation is that we sample radii for each vessel independently. While this choice allows us to investigate how geometry impacts hemodynamics, it overlooks the intimate connection between vessels in the system. When we sample independently, daughter radii can exceed their parent's, violating the morphometric condition. In the future, we will account for covariance enabling us to consider the link between parents and daughter vessels when sampling (McNeil, 2008).

We generate a relatively small number of segmentations for analysis in this study due to the time-consuming nature of manual segmentation of complex structures. We plan to create more segmentations of pulmonary and aortic vessels to test our method further. To do so, we will use automatic segmentation software, such as (Poletti et al., 2022), and compare this to the manual segmentations we have already completed. Another component is studying how changes in segmentation threshold and smoothing impact predictions. This was analyzed by Colebank et al. (2019) in excised mouse pulmonary artery data captured by microcomputed tomography images. Colebank's results had similar variation in flow and pressure compared to those reported in this study. Finally, it is important to factor in the noise in images. Animal dta are often obtained under anesthesia, while human data are obtained in the awake state. The latter may have higher impact on image resolution than analysis of radii extracted from the images.

## CONCLUSION

This study develops a robust framework to generate patient-specific labeled directed trees from medical images and obtain vessel measurements for use in fluid dynamics simulations while considering uncertainty. We find notable variation in the raw networks generated using VMTK, especially in the smallest vessels, when segmenting the same images multiple times despite using standard segmentation procedures and having the same user segment the images. To minimize this uncertainty, we develop a robust pipeline to prune networks, shift junction nodes to the center of the ostium, and determine vessel radius using change points. Our results emphasize the significance of post-processing and quantifying uncertainty when generating networks for medical applications. Current diagnostic procedures for many diseases require invasive protocol, including right-heart catheterization. This study demonstrates that careful analysis of image analysis data combined with CFD modeling has the potential to augment or mitigate the need for invasive studies via in-silico simulations. Despite the potential clinical advances CFD provides, medical conclusions should only be made by first assessing the level of uncertainty present in diagnostic information, fully understanding limitations and possible errors. In the absence of fixed geometric values of vessels due to scanning ambiguity, researchers must navigate uncertain measurements by developing effective methodologies, such as the one we have outlined here, for increasingly accurate analyses.

# Additional Information

## Data Availability

Software and code implementing the developed methodology can be found at

https://github.com/msolufse/CDG_NCSU/tree/master/VascularTreeFromImages.

## Competing Interests

The authors declare that they have no competing interests.

## Author Contributions

MAB and AMT designed the study, developed algorithms, performed hemodynamic simulations, and drafted the manuscript. DG, AJ, YL, ES, IS, and ZT contributed to algorithm development and image segmentation. JDW and CP provided aortic vasculature data and advised its use. DH provided statistical expertise and conceived the study. MSO conceived and coordinated the study.

## Funding

This work was supported by the National Science Foundation (grant numbers DGE-2137100, DMS-2051010), the National Security Agency (grant number H98230-23-1-0009), and the National Institute of Health (grant numbers NIH-HLBI RO1HL147590-02, NIH-5 T34 GM 1319475). Any opinions, findings, and conclusions expressed in this material are those of the authors and do not necessarily reflect the views of the NSF, NSA, or NIH.

## Acknowledgements

We thank Dr. Dan Lior, Baylor College of Medicine, Houston, TX for performing 4D-MRI registration and data extraction.